\documentclass[journal]{IEEEtran}
\ifCLASSINFOpdf
\else
\fi
\usepackage{amsmath,amsfonts}
\usepackage{algorithmic}
\usepackage{array}
\usepackage{textcomp}
\usepackage{stfloats}
\usepackage{url}
\usepackage{verbatim}
\usepackage{caption}
\usepackage{balance}
\usepackage{multicol}
\usepackage{tikz}
\usepackage[subrefformat=parens,labelformat=parens]{subcaption} 
\usepackage{tabularx} 
\usepackage{multirow}
\usepackage{graphicx}
\usepackage{pifont}

\usepackage{textcomp}  

\begin{document}
%
\title{Towards Intelligent Traffic Signaling in Dhaka City\\ Based on Vehicle Detection and\\ Congestion Optimization}
%
%
%


\author{Kazi Ababil Azam*,
        Hasan Masum*,
        Masfiqur Rahaman,
        and A. B. M. Alim Al Islam
\thanks{All authors were with the Department
of Computer Science and Engineering, Bangladesh University of Engineering and Technology, Dhaka, Bangladesh.}
\thanks{*Both authors contributed equally to this research.}%
}

\maketitle

\begin{abstract}
    The vehicular density in urbanizing cities of developing countries such as Dhaka, Bangladesh result in a lot of traffic congestion, causing poor on-road experiences. Traffic signaling is a key component in effective traffic management for such situations, but the advancements in intelligent traffic signaling have been exclusive to developed countries with structured traffic. The non-lane-based, heterogeneous traffic of Dhaka City requires a contextual approach. This study focuses on the development of an intelligent traffic signaling system feasible in the context of developing countries such as Bangladesh. We propose a pipeline leveraging Real Time Streaming Protocol (RTSP) feeds, a low resources system Raspberry Pi 4B processing, and a state of the art YOLO-based object detection model trained on the Non-lane-based and Heterogeneous Traffic (NHT-1071) dataset to detect and classify heterogeneous traffic. A multi-objective optimization algorithm, NSGA-II, then generates optimized signal timings, minimizing waiting time while maximizing vehicle throughput. We test our implementation in a five-road intersection at Palashi, Dhaka, demonstrating the potential to significantly improve traffic management in similar situations. The developed testbed paves the way for more contextual and effective Intelligent Traffic Signaling (ITS) solutions for developing areas with complicated traffic dynamics such as Dhaka City.
\end{abstract}


%
\IEEEpeerreviewmaketitle

\section{Introduction} \label{sec:introduction}
    Bangladesh, a developing nation in South East Asia, faces numerous urban challenges as it progresses economically. With one of the highest population densities among the cities of the world, its capital Dhaka also experiences a continuous increase in vehicles on its roads. This surge in traffic exerts immense pressure on the transportation infrastructure of the city. One of the most pressing issues in Dhaka is traffic congestion. 

\begin{figure}
\centering
  \begin{subfigure}[t]{.475\linewidth}
    \centering
    \includegraphics[width=\linewidth, height=0.55\textwidth]{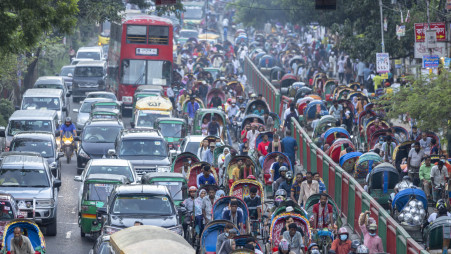}
    \caption{Heterogeneous traffic}
    \label{fig:nht1}
  \end{subfigure}
  \begin{subfigure}[t]{.475\linewidth}
    \centering
    \includegraphics[width=\linewidth, height=0.55\textwidth]{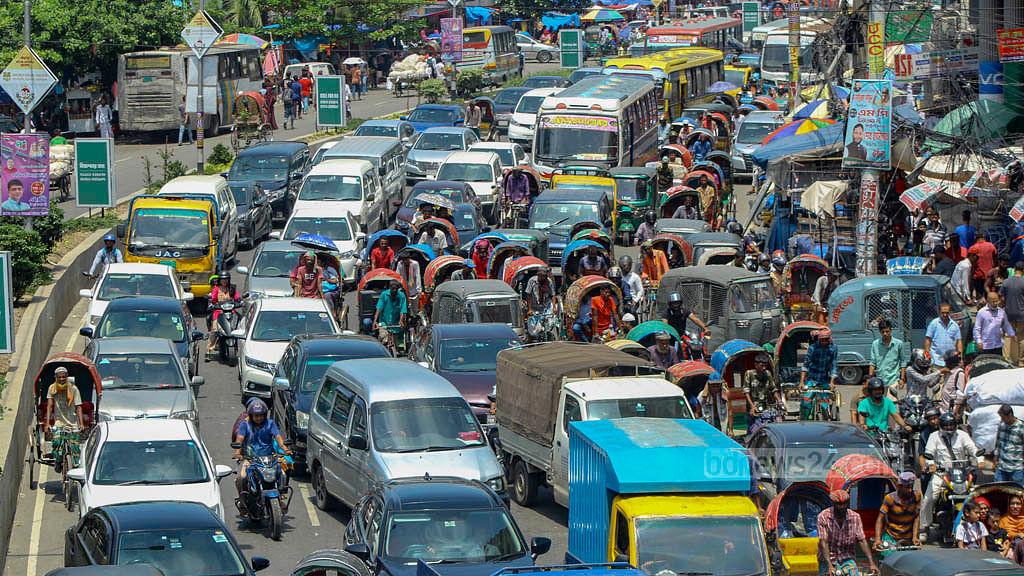}
    \caption{Lanes are rarely followed}
    \label{fig:nht2}
  \end{subfigure}
  \vspace{-0.25cm}
\caption{The non-lane-based, heterogeneous traffic in Dhaka}
\label{fig:nht}
\vspace{-0.25cm}
\end{figure}

Unlike the orderly traffic systems seen in developed nations, Dhaka's traffic is characterized by its non-lane-based and heterogeneous nature. The city's transportation system consists of a heterogeneous mix of vehicles that populate its roads, ranging from rickshaws to buses. 
The non-lane-based, heterogeneous traffic of Dhaka city as seen in Figure \ref{fig:nht} thus requires an efficient and intelligent traffic management system to control congestion and mitigate the terrible on-road experience in the city.

\begin{figure}
\centering
  \begin{subfigure}[t]{.475\linewidth}
    \centering
    \includegraphics[width=\linewidth, height=0.55\textwidth]{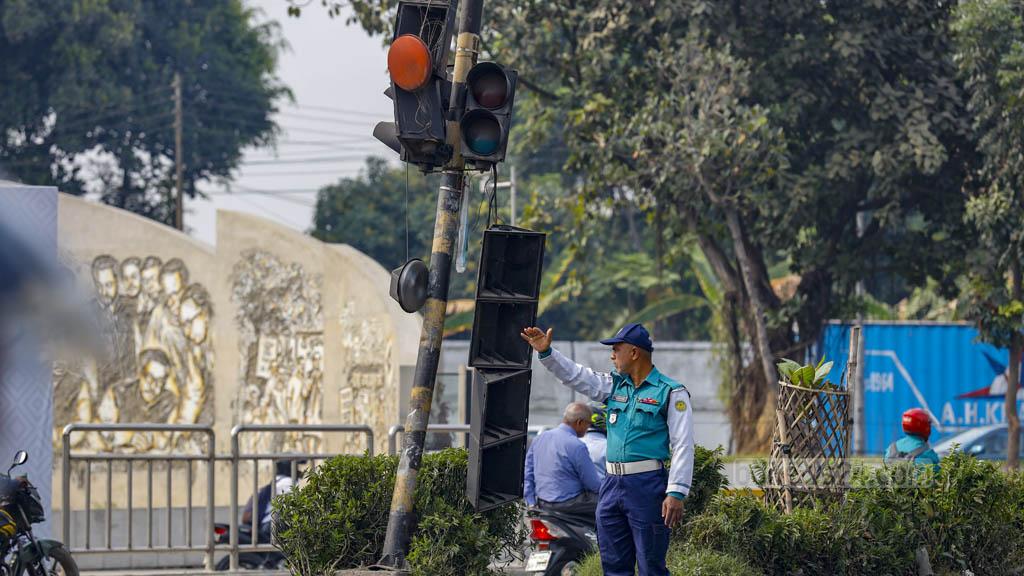}
    \caption{Manual signaling}
    \label{fig:traffic-police}
  \end{subfigure}
  \begin{subfigure}[t]{.475\linewidth}
    \centering
    \includegraphics[width=\linewidth, height=0.55\textwidth]{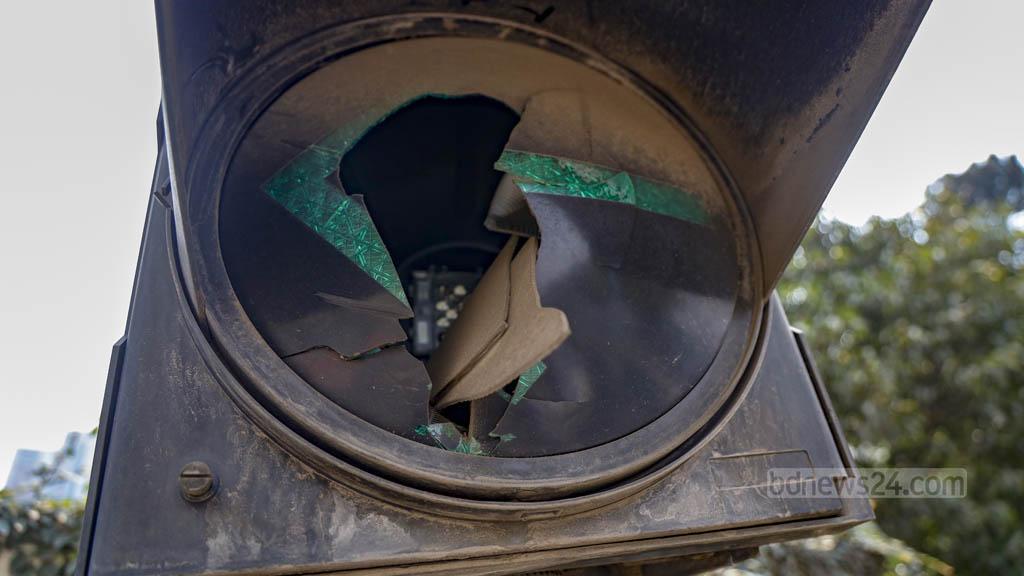}
    \caption{Inoperative traffic signals}
    \label{fig:inoperative}
  \end{subfigure}
\vspace{-0.25cm}
\caption{Condition of traffic signaling  in Dhaka}
\label{fig:condn-traffic-signaling}
\vspace{-0.5cm}
\end{figure}

Traffic signaling is one of the major components of the traffic management infrastructure of any city or country. Effective traffic signal management helps to mitigate congestion, reduce travel times, and enhance road safety. Moreover, traffic signaling plays a pivotal role in reducing fuel consumption and emissions by minimizing idle times and ensuring smoother traffic flow. But unfortunately, the traffic signaling system in Dhaka city is in terrible shape. 
Almost no automated traffic light system is actively working in Dhaka city, with the only two exceptions being the Gulshan and Dhaka Cantonment areas.  Most of the critical junctions in Dhaka are managed by the traffic police using their hand-generated signals, as shown in Figure \ref{fig:traffic-police}. They coordinate between themselves using some communication devices, such as wireless walkie-talkies and whistles. 
Figure \ref{fig:inoperative} shows such a traffic light.

Various intelligent solutions have been proposed to address traffic congestion and management challenges. Nonetheless, most image-based traffic signaling technologies, for instance, are crafted for the structured, lane-based transportation systems seen in developed countries \cite{chen2020future, yogheshwaran2020iot, liu2020thresholds, beg2021uav, zhou2021construction}. On the other hand, the traffic flow in Dhaka does not adhere to lanes. This leads to significant issues, causing the conventional image-based models designed with developed nations in mind less effective. Feasibility of deployment and operation is also an issue in the context of developing countries such as Bangladesh. The advanced technological implementations in developed countries are financially infeasible for developing countries to integrate into their existing infrastructure. Thus, there is a pressing requirement to design a novel image-based traffic signaling approach specifically tailored to the unique requirements of busy cities in developing countries. The system should be based on state-of-the-art technology to maximize accuracy, with the flexibility of implementation in a cost-effective manner.

In response to these requirements, we propose an intelligent traffic signaling system fitting to the requirements of Dhaka's traffic landscape. Our solution leverages Real Time Streaming Protocol (RTSP) feeds to capture live traffic images at intersections. This data is processed using a Raspberry Pi 4B, a low-resource computing platform, which makes our system both feasible and scalable for deployment. Central to our approach is the use of a state-of-the-art YOLO-based object detection model, trained specifically on the Non-lane-based and Heterogeneous Traffic (NHT-1071) dataset \cite{mrahman23}. We utilize the Non-dominated Sorting Genetic Algorithm II (NSGAII) \cite{nsgaii} to optimize traffic signal timings, balancing the objectives of minimizing waiting times and maximizing vehicle throughput, as evaluated in \cite{mrahman23}. We test our system at a five-road intersection in Palashi, Dhaka, demonstrating its potential to alleviate traffic congestion and improve overall traffic flow. Overall, we make the following contributions in this study,

\begin{itemize}
    \item We propose a novel system architecture for traffic signal scheduling that incorporates successive implementations of vehicle detection and multi-objective optimization algorithms in limited-resource environments, specifically focused on the traffic scenario of Dhaka city.
    \item We benchmark the NHT-1071 \cite{mrahman23} dataset on a Raspberry Pi using state-of-the-art detection models, evaluating performance and resource efficiency.
    \item We develop a scalable pipeline that incorporates the vehicle detection model and multi-objective optimization algorithm NSGA-II \cite{nsgaii}, ensuring robust and adaptable traffic management for various kinds of implementation.
    \item We implement the proposed system architecture in a cost-effective manner and test the system in both simulated environments and actual on-road conditions, validating its feasibility, reliability and efficiency in real-life traffic scenarios.
\end{itemize}
\section{Background and Related Studies} \label{sec:background}

Traffic signal scheduling is an important  transportation systems. Scheduling algorithms affect metrics such as average speed of the vehicles, time and volume of congestion, and flow rate of traffic which depict the on-road experience of a traffic scenario.
Common scheduling techniques include fixed-time traffic signals, actuated traffic signals, and adaptive traffic signals. Fixed-time traffic signals operate on a predetermined cycle, with each phase (e.g., green, yellow, red) lasting for a fixed duration based on historical traffic patterns. Actuated traffic signals use sensor systems to detect the presence or absence of vehicles, allowing signals to change based on real-time traffic conditions and demand. Adaptive traffic signals employ advanced algorithms and real-time data to adjust signal timing dynamically based on current traffic conditions.

Recently, automated traffic signal scheduling systems have become increasingly prevalent in managing the flow of vehicles. Usually, the duration that each signal stays green for a particular lane or direction is predetermined for all directions. This fixed-time approach often struggles to accommodate heavy traffic in specific lanes, leading to inefficient management. Consequently, numerous innovative technologies have been introduced to enhance the automation of traffic signal systems and address these challenges \cite{saha2020automated, tajalli2020network, al2019smart, waddell2020characterizing, yao2020reducing, chen2021rhythmic, almawgani2018design}. 

There are plenty of recent studies on intelligent traffic signaling which have worked with vehicle detection and multi-objective optimization. The studies propose various reinforcement learning (RL) and artificial intelligence (AI) based models for traffic signal control, primarily focusing on improving traffic flow and reducing congestion. Zhang et al. presented an RL-based system effective under low detection rates but required costly infrastructure and considered only motorized traffic \cite{zhang2020}. Mandal et al. designed an AI-enabled monitoring system based on New York freeway images but was limited to monitoring motorized traffic \cite{mandal2020}. Joo et al. proposed an algorithm using deep Q-networks for traffic signal optimization, but it was tested only in developed countries with structured traffic \cite{joo2022}. Most recently Leal et al. utilized NSGA-II for traffic light optimization but did not expand on deployment \cite{leal2023}. The dataset and study was based on Brazil, the traffic of which also is more structured than Bangladesh. Rai et al. developed an IoT-based intelligent signaling system which was tested only in simulations of basic four-way intersections \cite{rai2023}. 
Other studies similarly focused on structured, motorized traffic in developed countries, limiting the scope to homogeneous traffic \cite{jahan2020analyzing, maungmai2019vehicle, liu2018urban, tan2017optimization, nguyen2016discovering}. 
\begin{figure}[t]
    \centering
    \includegraphics[width=0.45\textwidth]{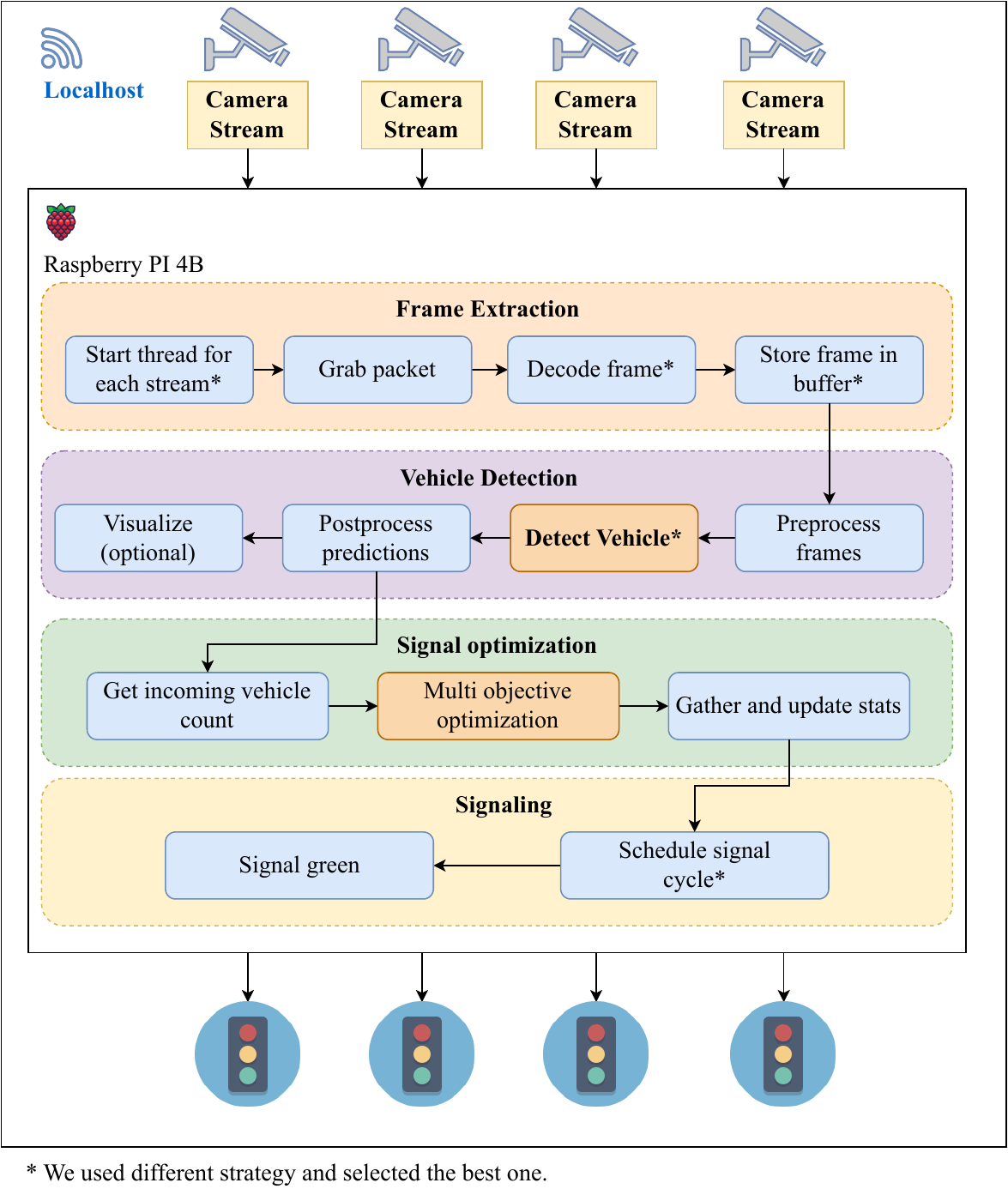}
    \caption{System architecture of our proposed system}
    \label{fig:sysarch}
    \vspace{-0.5cm}
\end{figure}

Toha et al. proposed DhakaNet \cite{dhakanet}, a limited-resource deep learning architecture, DhakaNet, for real-time vehicle detection from on-road traffic images by modifying the then state-of-the-art object detection model, YOLOv5 \cite{yolov5}. The architecture was designed to readily detect vehicles in low-resource environments such as Raspberry Pi. The motivation of the study aligns with the context of non-lane-based heterogeneous traffic of Dhaka. However, the study did not propose any contextually-appropriate image dataset. It did not elaborate on the optimization or implementation of any system to work with DhakaNet. The base model has been further updated with the astounding advancements in object detection, with YOLOv8 \cite{yolov8_ultralytics}, YOLOv9 \cite{wang2024yolov9}, YOLOv10 \cite{THU-MIGyolov10}, and most recently YOLOv11.

The most recent advancement in the context of non-lane-based heterogeneous traffic analysis and traffic signaling has been the presentation of an image dataset suited for Dhaka traffic to train and test object detection models, and a methodology to integrate mentioned models with a multi-objective optimization algorithm to generate signal cycles \cite{mrahman23}. The dataset is meticulously prepared with 1071 traffic images from the roads of different areas in Dhaka city, which we use in our study. 

\section{System Architecture} \label{sec:system-arch}

Figure \ref{fig:sysarch} illustrates our proposed intelligent traffic signaling system architecture, which addresses the gaps identified in Section \ref{sec:background}. The whole system runs on a local network consisting of a Raspberry Pi and IP cameras. These cameras send RTSP streams over a wireless network to the Raspberry Pi, which processes these feeds in real-time.

Stream processing consists of three main stages: frame extraction, vehicle detection, and signal optimization. In the frame extraction stage, one thread is initiated for each camera stream to grab packets. The packets are conditionally decoded when needed and stored in a buffer of size one. We use an eager strategy to fill the buffer, meaning that the oldest frames are dropped, and only the latest frame is kept. The vehicle detection stage then processes this latest frame in another thread. Finally, the signal optimization stage uses the incoming vehicle counts from vehicle detection frames to generate optimized traffic signal timings. This minimizes waiting time for vehicles and maximizes vehicle throughput. The optimized signal cycles are then scheduled and implemented to manage traffic flow effectively. This architecture is designed to operate efficiently on low-resource hardware like the Raspberry Pi, making it suitable for deployment in resource-constrained environments typical of developing countries.
\begin{figure}[t]
    \centering
    \includegraphics[width=\linewidth]{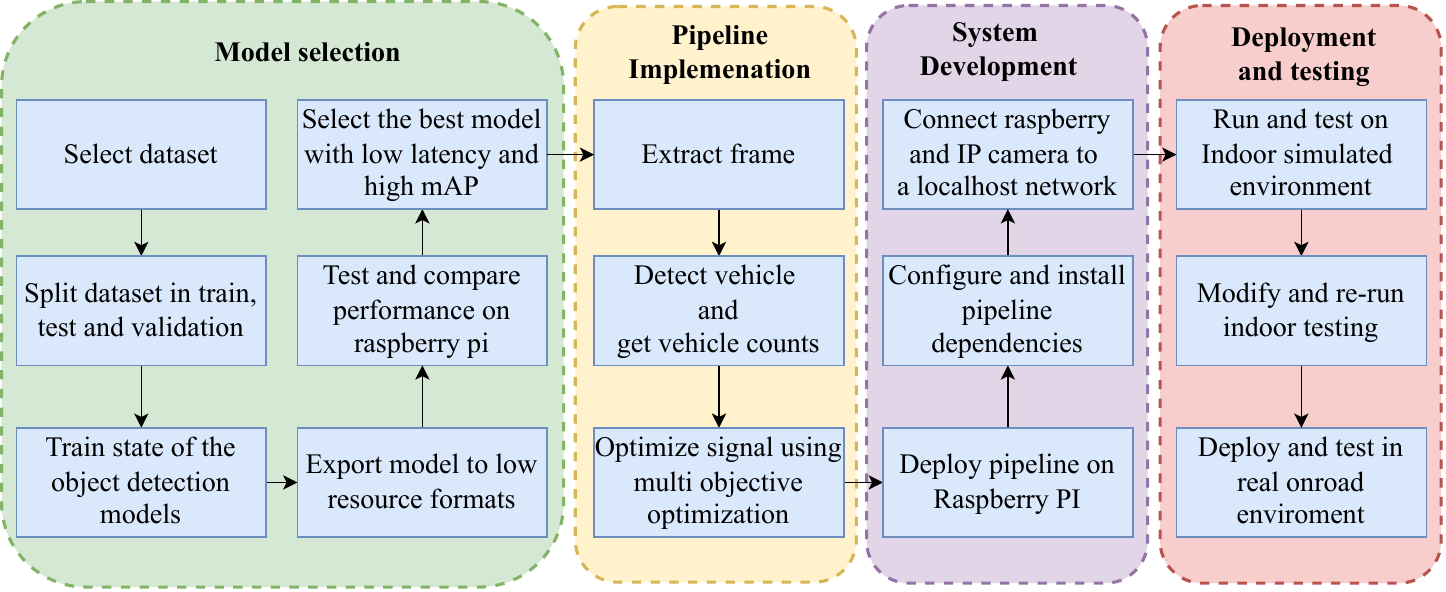}
    \caption{Methodology of our study}
    \label{fig:methodology}
    \vspace{-0.5cm}
\end{figure}
\vspace{-0.2cm}
\subsection{Frame Extraction}
\subsubsection{Real-time Traffic Data Acquisition} 
The system retrieves real-time traffic video feeds using the Real-Time Streaming Protocol (RTSP). This protocol enables efficient transmission of live video streams from IP cameras installed at the intersection.

\subsubsection{Multi-Threaded Frame Extraction with PyAV\hspace{-0.15em}} 
The PyAV library is employed to efficiently extract video frames from the incoming RTSP streams. A multi-threading approach is implemented to parallelize the frame extraction process. This significantly improves performance by utilizing multiple CPU cores on the Raspberry Pi, ensuring smooth handling of high-resolution video streams. Concurrent frame extraction from multiple cameras allows for real-time processing without delays. Improved responsiveness to traffic fluctuations by enabling faster analysis of the latest video frames.

\vspace{-0.2cm}
\subsection{Vehicle Detection}
Similar to frame extraction, multi-threading is implemented to expedite object detection within the extracted video frames. Each worker thread in the pool processes a frame using the chosen model, identifying and classifying objects present in the scene. Parallelized inference significantly reduces processing time, enabling near real-time analysis of traffic conditions.
Efficient utilization of the Raspberry Pi's processing power by distributing the workload across multiple cores.

\vspace{-0.2cm}
\subsection{Signal optimization}
\subsubsection{Traffic Data Analysis} 
The pipeline accumulates object detection results (number and type of vehicles) over a predefined time window. This data provides a comprehensive understanding of the current traffic volume and composition at the intersection.

\subsubsection{NSGA-II for Multi-Objective Optimization}
The Non-dominated Sorting Genetic Algorithm II (NSGA-II) \cite{nsgaii} is employed as the optimization algorithm \cite{mrahman23}. NSGA-II is well-suited for multi-objective optimization problems, making it ideal for this scenario where we aim to (1) minimize waiting time for vehicles at the intersection, and (2) maximize vehicle throughput (number of vehicles passing through the intersection per unit time).
NSGA-II's core features of handling conflicting objectives like those mentioned above, maintaining solution diversity, and leveraging real-time data make it an excellent fit for optimizing traffic signal timing to minimize wait times and maximize vehicle throughput.

\section{System Design} \label{sec:system-design}

This section details the methodological approach taken to develop an intelligent traffic signaling system suitable for the context of Dhaka City, Bangladesh. Our study can be broken down into model selection, process pipeline implementation, system development, deployment and testing, as shown in Figure \ref{fig:methodology}.

\subsection{Object Detection Model Selection}

The first step in building our system is model selection. 
Accurate detection and classification of vehicles are crucial for the system's functionality. This section outlines the process of selecting an object detection model suitable for Dhaka's non-lane-based and heterogeneous traffic conditions.

\subsubsection{Dataset Description}
\begin{figure}[!tbp]
    \centering
  \begin{subfigure}[t]{0.23\linewidth}
    \centering
    \includegraphics[width=0.9\linewidth]{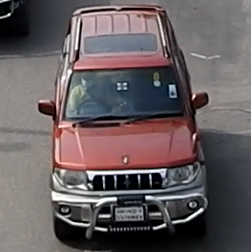}
    \caption{Motorized incoming}
  \end{subfigure}
\hfill
  \begin{subfigure}[t]{0.23\linewidth}
    \centering
    \includegraphics[width=0.9\linewidth]{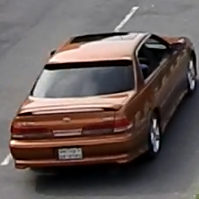}
    \caption{Motorized outgoing}
  \end{subfigure}
\hfill
  \begin{subfigure}[t]{0.23\linewidth}
    \centering
    \includegraphics[width=0.9\linewidth]{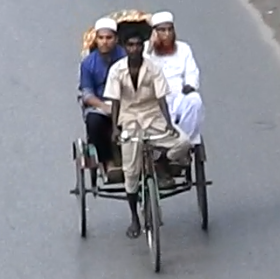}
    \caption{Non motorized incoming}
  \end{subfigure}
\hfill
  \begin{subfigure}[t]{0.23\linewidth}
    \centering
    \includegraphics[width=0.9\linewidth]{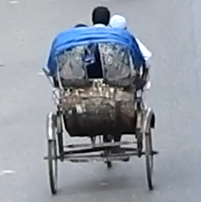}
    \caption{Non motorized outgoing}
  \end{subfigure}
  \vspace{-0.25cm}
\caption{Class labels of NHT-1071 dataset}
\label{fig:data_labels}
\vspace{-0.55cm}
\end{figure}

The vehicle detection system relies on a comprehensive dataset known as Non-lane-based and Heterogeneous Traffic (NHT-1071). This dataset encompasses 1071 images capturing the diverse traffic conditions of Dhaka City, including variations in weather and illumination. The dataset contains approximately 30,000 objects labeled across four distinct classes (Figure \ref{fig:data_labels}), reflecting the variety of vehicles present on Dhaka's roads. 

NHT-1071 classifies vehicles into the following categories: 
(1) Motorized vehicles entering the intersection (motorized in), 
(2) Non-motorized vehicles entering the intersection (non-motorized in), 
(3) Motorized vehicles exiting the intersection (motorized out), 
and (4) Non-motorized vehicles exiting the intersection (non-motorized out).

Dhaka City experiences non-lane-based traffic. Consequently, existing image-based datasets designed for developed countries may not perform well. These datasets often only focus on the type of vehicle, neglecting the direction of movement. NHT-1071, specifically created for Dhaka's non-lane-based and heterogeneous traffic, considers vehicle direction. This makes it ideal for our context.

\subsubsection{Lightweight Object Detection Model}

We trained state of the art CNN based YOLO series of real-time object detectors -YOLOv8\cite{yolov8_ultralytics}, YOLOv9\cite{wang2024yolov9}, YOLOv10\cite{THU-MIGyolov10} and Real-Time Detection Transformer (RT-DETR) \cite{lv2023detrs} on NHT-1071 dataset.

Among those models, we select the lightweight versions, such as YOLOv8 medium, YOLOv8 small, YOLOv8 nano, YOLOv9 medium, YOLOv9 small, YOLOv9 tiny, YOLOv8 medium, YOLOv10 small, YOLOv10 nano and RTDETR light for the experimental evaluation.

\subsubsection{Evaluation metrics}
We use mean average precision (mAP) at intersection over union (IOU) threshold of 50 (mAP@50) and mAP at IOU threshold of 50-95 (mAP@50-95) for evaluating model's performance across different levels of detection difficulty. 

\subsubsection{Model selection criteria}
We train and evaluate the candidate models (YOLOv8, YOLOv9, YOLOv10, and RTDet) on the NHT-1071 dataset. The model achieving the best balance between accuracy (high mAP@50 and mAP@50-90), size (fewer parameters), and low inference time will be chosen for deployment in the intelligent traffic signaling system. This approach prioritizes both accurate object detection and real-time performance, ensuring the system's effectiveness in the resource-constrained environment of the Raspberry Pi. 

\vspace{-0.2cm}
\subsection{Multi objective optimization}
\begin{figure}[!t]
    \centering
    \includegraphics[width=0.55\linewidth]{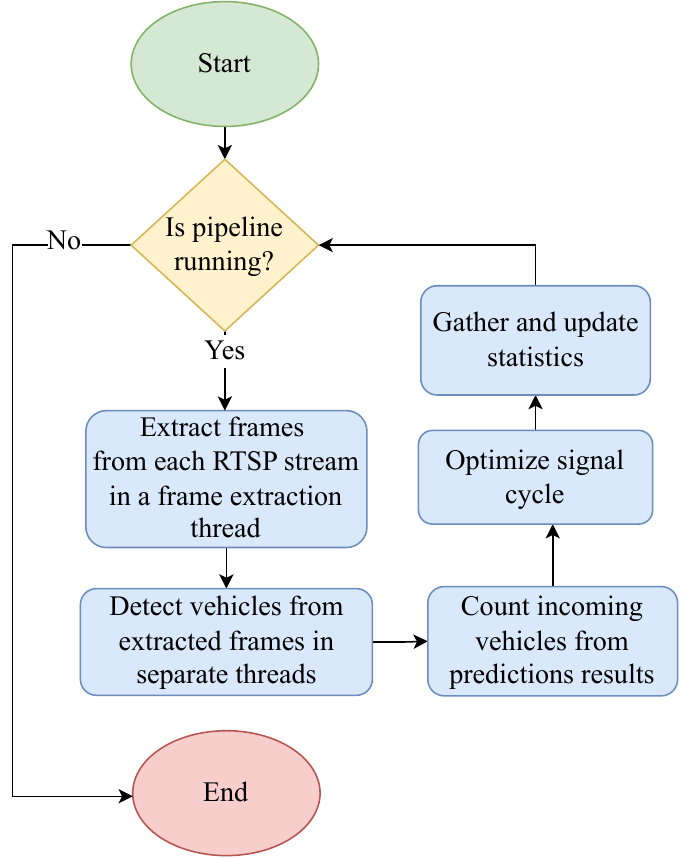}
    \vspace{-0.25cm}
    \caption{Flow chart of the entire pipeline}
    \label{fig:methodoloy:pipeline:flowchart}
    \vspace{-0.55cm}
\end{figure}

For traffic signal optimization in the roads of Dhaka, we need to reduce traffic congestion volume and reduce waiting time of the vehicles as well. These two are conflicting objectives which makes our problem suitable for multi-objective optimization. We use objective functions that are designed to achieve these goals in a recent work \cite{mrahman23}. 
Given the count of motorized and non-motorized vehicles for each link. The first objective function of the algorithm calculates the total updated congestion by summing the updated vehicle counts for each link after they uses their assigned green time.

Let L be the number of links in the intersection. Let $C_m(i)$ and $C_{nm}(i)$ signify the count of motorized and non-motorized vehicles on link $i$ respectively. Then first objective function is given by,  
\vspace{-0.4cm}
\begin{equation}
\label{eq:obj1}
f_1 = \sum_{i=1}^{L} (C_m(i) + C_{nm}(i))
\end{equation}

The second objective function of the algorithm is to calculate the waiting time for vehicles on links with red signals, using the red signal time as an approximate waiting time.

Let $R_t(i)$ be the total red signal time for link $i$ on that signal cycle.  Then the second objective function is given by,
\vspace{-0.4cm}

\begin{equation}
\label{eq:obj2}
f_2 = \sum_{i=1}^{L} R_t(i)
\end{equation}

Both equations are evaluated after a complete signal cycle for all links at an intersection. Equation \ref{eq:obj1} provides the total number of vehicles, both motorized and non-motorized, stuck in congestion. Equation \ref{eq:obj2} gives the total waiting time for vehicles at red signals. Minimizing these values are our priority. \cite{mrahman23}
\section{Implementation} \label{sec:implementation}
This section details the development of the intelligent traffic signaling process pipeline, which processes real-time traffic data, performs object detection, and optimizes signal timings. Figure \ref{fig:methodoloy:pipeline:flowchart} illustrates the flowchart of the pipeline.


\vspace{-0.2cm}
\subsection{Frame Extraction with Multi-Threading}
\begin{figure}[!t]
  \begin{subfigure}[t]{0.5\linewidth}
    \centering
    \includegraphics[width=\linewidth]{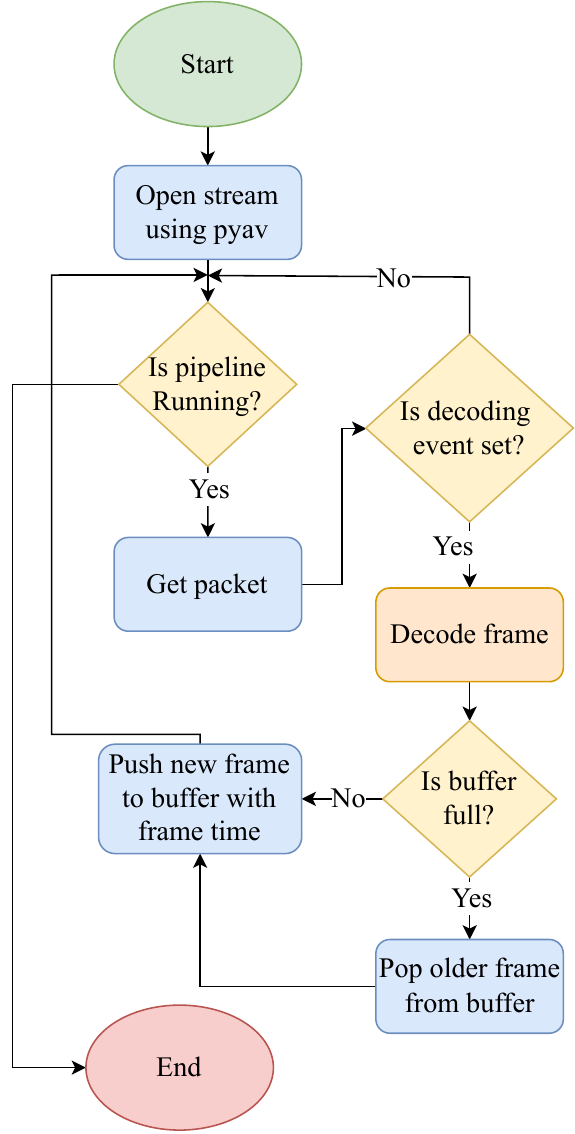}
    \caption{Frame extraction thread}
    \label{fig:pipeline:frame_extraction}
  \end{subfigure}
\hfill
  \begin{subfigure}[t]{0.48\linewidth}
    \centering
    \includegraphics[width=\linewidth]{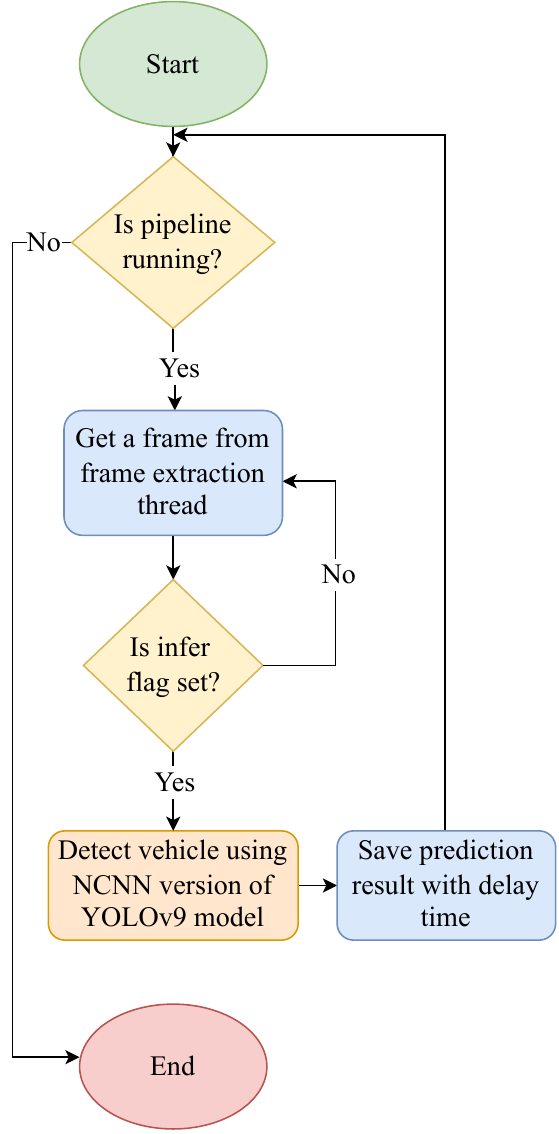}
    \caption{Inference thread}
    \label{fig:pipeline:inference_thread}
  \end{subfigure}
  \vspace{-0.25cm}
  \caption{Flow chart of two concurrent processes for real-time object detection. One process extracts frames from a video stream, while the other process analyzes the frames to detect objects.}
  \vspace{-0.25cm}
\end{figure}

Figure \ref{fig:pipeline:frame_extraction} shows the flowchart of the frame extraction thread. The pipeline establishes connections to each Ezviz IP camera using their respective RTSP URLs.
PyAV's decoding functionalities are leveraged to extract individual video frames from the received streams.
A thread pool is created to manage multiple worker threads. Each thread is assigned the task of decoding a frame from a specific camera stream. Extracted frames are stored in a buffer or queue to be readily accessible for subsequent processing stages.

\vspace{-0.2cm}
\subsection{Frame Inference with Multi-Threading}

Figure \ref{fig:pipeline:inference_thread} shows the flowchart of the inference thread. The pre-trained object detection model is loaded onto the Raspberry Pi.
The thread pool established for frame extraction is reused for inference. Each thread receives a frame from the
buffer and performs object detection using the loaded model.
The model outputs bounding boxes and class labels for the detected objects within the frame.
This information is stored alongside the corresponding frame for further processing.

\vspace{-0.2cm}
\subsection{Signal Optimization using Multi-objective optimization}
\begin{figure}[!t]
    \centering
    \includegraphics[width=0.5\linewidth]{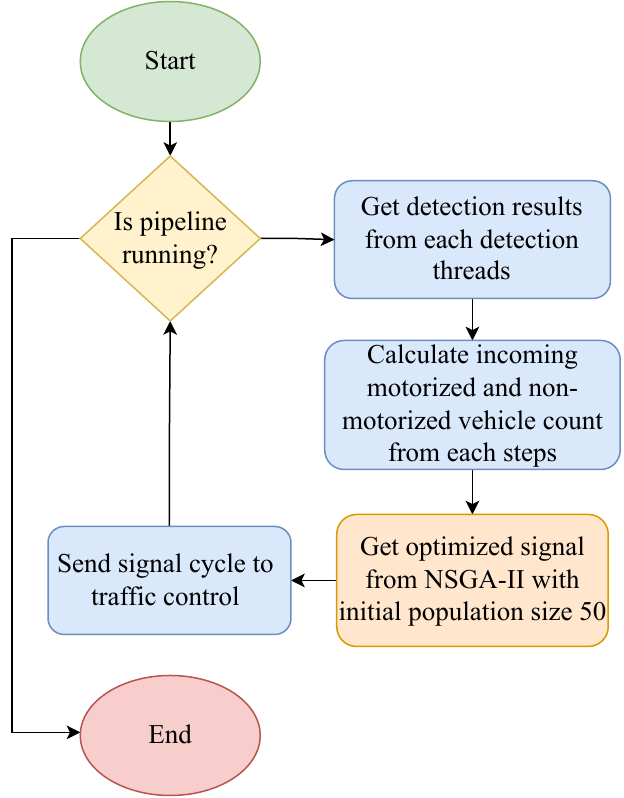}
    \vspace{-0.25cm}
    \caption{Flow chart of the signal optimization.}
    \label{fig:methodoloy:pipeline:signaling}
    \vspace{-0.25cm}
\end{figure}

Figure \ref{fig:methodoloy:pipeline:signaling} shows the flowchart of how we use NSGA-II in our implementation. The pipeline translates the collected traffic data (vehicle counts and types) into a format suitable for NSGA-II.
NSGA-II is configured with two objectives (minimizing waiting time and maximizing throughput) and appropriate constraints (e.g., minimum green time for each direction).
The algorithm iteratively evolves a population of candidate signal timing solutions, evaluating each solution based on its performance in achieving both objectives simultaneously.
The final output of NSGA-II is the optimized traffic signal timing plan that best balances minimizing waiting time and maximizing throughput for the current traffic conditions. 

\section{Experimentation and Evaluation} \label{sec:experimentation-and-evaluation}
    We evaluate our system and discuss experiments conducted to assess its reliability and feasibility.


\vspace{-0.2cm}
\subsection{Model Selection}
We evaluate state of the art object detection models and select a model to implement in our system.
\begin{table*}[h]
\caption{Model Performance Comparison in GPU}
\centering
\resizebox{0.95\textwidth}{!}{
\begin{tabular}{cccccccccc}
\hline
\textbf{Model}    & \textbf{\begin{tabular}[c]{@{}c@{}}mAP\\ @50\end{tabular}} & \textbf{\begin{tabular}[c]{@{}c@{}}mAP\\ @50-95\end{tabular}} & \textbf{Precision} & \textbf{Recall} & \textbf{\begin{tabular}[c]{@{}c@{}}Params\\ (Million)\end{tabular}} & \textbf{\begin{tabular}[c]{@{}c@{}}Preprocess \\ (ms)\end{tabular}} & \textbf{\begin{tabular}[c]{@{}c@{}}Inference \\ (ms)\end{tabular}} & \textbf{\begin{tabular}[c]{@{}c@{}}Postprocess\\ (ms)\end{tabular}} & \textbf{\begin{tabular}[c]{@{}c@{}}Latency\\ (ms)\end{tabular}} \\ \hline
YOLOv8n           & 0.621                                                      & 0.353                                                         & 0.695              & 0.592           & 3.01                                                                & 1.6                                                                 & 7.2                                                                & 21.3                                                                & 30.1                                                            \\
YOLOv8s           & 0.696                                                      & 0.414                                                         & 0.737              & 0.664           & 11.13                                                               & 1.4                                                                 & 13                                                                 & 22.3                                                                & 36.7                                                            \\
YOLOv8m           & 0.72                                                       & 0.422                                                         & 0.742              & 0.698           & 25.84                                                               & 1.1                                                                 & 24.3                                                               & 22.4                                                                & 47.8                                                            \\
YOLOv9t           & 0.625                                                      & 0.361                                                         & 0.696              & 0.588           & 1.97                                                                & 1.4                                                                 & 12                                                                 & 23.7                                                                & 37.1                                                            \\
YOLOv9s           & 0.689                                                      & 0.41                                                          & 0.743              & 0.64            & 7.17                                                                & 1.4                                                                 & 29.1                                                               & 25.1                                                                & 55.6                                                            \\
YOLOv9m           & 0.738                                                      & 0.433                                                         & 0.755              & 0.728           & 20.02                                                               & 1.4                                                                 & 31.8                                                               & 24.3                                                                & 57.5                                                            \\
\textbf{YOLOv10n} & \textbf{0.59}                                              & \textbf{0.336}                                                & \textbf{0.659}     & \textbf{0.552}  & \textbf{2.69}                                                       & \textbf{1.5}                                                        & \textbf{13.7}                                                      & \textbf{1.7}                                                        & \textbf{16.9}                                                   \\
\textbf{YOLOv10s} & \textbf{0.674}                                             & \textbf{0.396}                                                & \textbf{0.736}     & \textbf{0.631}  & \textbf{8.04}                                                       & \textbf{1}                                                          & \textbf{17.5}                                                      & \textbf{2.9}                                                        & \textbf{21.4}                                                   \\
\textbf{YOLOv10m} & \textbf{0.699}                                             & \textbf{0.41}                                                 & \textbf{0.715}     & \textbf{0.685}  & \textbf{16.45}                                                      & \textbf{1.6}                                                        & \textbf{22.8}                                                      & \textbf{1.9}                                                        & \textbf{26.3}                                                   \\
YOLOv11n          & 0.636                                                      & 0.363                                                         & 0.686              & 0.584           & 2.58                                                                & 1.4                                                                 & 13.4                                                               & 11.0                                                                & 25.8                                                            \\
YOLOv11s          & 0.696                                                      & 0.413                                                         & 0.686              & 0.686           & 9.41                                                                & 0.9                                                                 & 19.9                                                               & 15.6                                                                & 36.4                                                            \\
YOLOv11m          & 0.731                                                      & 0.437                                                         & 0.71               & 0.715           & 20.03                                                               & 1.5                                                                 & 30.2                                                               & 15.6                                                                & 47.3                                                            \\
RT-DETR           & 0.79                                                       & 0.444                                                         & 0.824              & 0.719           & 31.99                                                               & 0.4                                                                 & 43.9                                                               & 3                                                                   & 47.3                                                            \\ \hline
                                                    
\end{tabular}
}
\label{tab:model-performance}
\end{table*}

\subsubsection{Training and Evaluation Setups}
At first we train and evaluate the state-of-the-art models on the NHT-1071 dataset \cite{mrahman23}. The experimental setup for training and GPU evaluation was conducted in a Kaggle environment running Linux OS. The Python version used was 3.10.13, and the Ultralytics library version was 8.2.38. The experiments utilized PyTorch 2.1.2 and a Tesla P100-PCIE-16GB GPU with 16276MiB of memory. CUDA version 12.2 was employed for GPU acceleration.. We trained the each model for 50 epochs with batch size 4 and image size 768.

But we require the models to perform well on CPU environments, and there are frameworks such as ONNX ans NCNN to export models for optimization in CPU and edge devices. Table \ref{tab:export-support} shows the exporting support for the models we opted to evaluate. 

We then evaluate the models on a Windows Computer CPU. The experimental setup for the Windows CPU evaluation was conducted on a PC running Windows 10. The Python version used was 3.9.5, with the Ultralytics library at version 8.2.38 and PyTorch 2.1.2+cpu. No GPU was utilized, and the CPU was an Intel Core i5-8300H at 2.30GHz. 

Finally we evaluate the models on our working environment, Raspberry Pi CPU on setup. The experimental setup for the Raspberry Pi 4B evaluation was conducted on an arch64 Linux operating system. The Python version used was 3.11.2, with the Ultralytics library at version 8.2.38 and PyTorch 2.3.1+cpu. No GPU was utilized, and the CPU was a Cortex-A72.

\begin{table}[t]
\caption{Exporting support of different models}
\vspace{-0.2cm}
\centering
\begin{tabular}{|c|c|c|}
\hline
\textbf{Model Name} & \textbf{ONNX} & \textbf{NCNN} \\ \hline
YOLOv8              & Yes           & Yes           \\ \hline
YOLOv9              & Yes           & Yes           \\ \hline
YOLOv10             & Yes           & No            \\ \hline
RTDETR              & No            & No            \\ \hline
\end{tabular}
\label{tab:export-support}
\vspace{-0.2cm}
\end{table}

\subsubsection{Performance and Selection}

After training and evaluating, we find that the YOLOv10 models perform the best among the tested models, as shown in Table \ref{tab:model-performance}. This is due to the low post processing time, improving the overall inference time by a lot.

We then evaluate the models on a Windows Computer CPU and finally our working environment, Raspberry Pi CPU, the results of which are shown in Table \ref{tab:model-perf-pc} and \ref{tab:model-perf-pi} respectively. We select the NCNN \cite{ncnn} exported model of YOLOv9, as it exceeds all other models in terms of both inference time and mean average precision.

\begin{table*}[h]
\caption{Model Performance Comparison in Windows PC CPU}
\resizebox{0.95\textwidth}{!}{
\begin{tabular}{cccccccccc}
\hline
\textbf{Model}                    & \textbf{Format} & \textbf{\begin{tabular}[c]{@{}c@{}}mAP\\ @50\end{tabular}} & \textbf{\begin{tabular}[c]{@{}c@{}}mAP\\ @50-95\end{tabular}} & \textbf{Precision} & \textbf{Recall} & \textbf{Params}  & \textbf{\begin{tabular}[c]{@{}c@{}}Preprocess\\ (ms)\end{tabular}} & \textbf{\begin{tabular}[c]{@{}c@{}}Inference\\ (ms)\end{tabular}} & \textbf{\begin{tabular}[c]{@{}c@{}}Postprocess\\ (ms)\end{tabular}} \\ \hline
\multirow{3}{*}{YOLOv8n}          & pytorch         & 0.692                                                      & 0.39                                                          & 0.752              & 0.615           &                  & 7.7                                                                & 401.7                                                             & 1.7                                                                 \\
                                  & onnx            & 0.685                                                      & 0.385                                                         & 0.742              & 0.618           & 3006428          & 10                                                                 & 123.1                                                             & 2.7                                                                 \\
                                  & ncnn            & 0.685                                                      & 0.385                                                         & 0.742              & 0.618           &                  & 11.8                                                               & 171.1                                                             & 3.4                                                                 \\ \hline
\multirow{3}{*}{\textbf{YOLOv9t}} & pytorch         & 0.686                                                      & 0.385                                                         & 0.747              & 0.621           & \textbf{}        & 5.6                                                                & 394.9                                                             & 1.6                                                                 \\
                                  & \textbf{onnx}   & \textbf{0.687}                                             & \textbf{0.383}                                                & \textbf{0.746}     & \textbf{0.621}  & \textbf{1971564} & \textbf{8.9}                                                       & \textbf{111.6}                                                    & \textbf{2.6}                                                        \\
                                  & ncnn            & 0.687                                                      & 0.383                                                         & 0.746              & 0.621           &                  & 10.8                                                               & 170.2                                                             & 3                                                                   \\ \hline
\multirow{2}{*}{YOLOv10n}         & pytorch         & 0.659                                                      & 0.367                                                         & 0.684              & 0.598           & 2695976          & 7.8                                                                & 441.3                                                             & 0.1                                                                 \\
                                  & onnx            & 0.657                                                      & 0.366                                                         & 0.695              & 0.597           &                  & 10.4                                                               & 116.3                                                             & 0.1                                                                 \\ \hline
\end{tabular}
}
\label{tab:model-perf-pc}
\end{table*}

\begin{table*}[!h]%
\centering
\caption{Model Performance Comparison in Raspberry Pi CPU}
\resizebox{0.95\linewidth}{!}{
\begin{tabular}{ccccccccccc}
\hline
\textbf{Model}                    & \textbf{Format} & \textbf{\begin{tabular}[c]{@{}c@{}}mAP\\ @50\end{tabular}} & \textbf{\begin{tabular}[c]{@{}c@{}}mAP\\ @50-95\end{tabular}} & \textbf{Precision} & \textbf{Recall} & \textbf{Params}                   & \textbf{\begin{tabular}[c]{@{}c@{}}Preprocess\\ (ms)\end{tabular}} & \textbf{\begin{tabular}[c]{@{}c@{}}Inference\\ (ms)\end{tabular}} & \textbf{\begin{tabular}[c]{@{}c@{}}Postprocess\\ (ms)\end{tabular}} & \textbf{\begin{tabular}[c]{@{}c@{}}Latency\\ (ms)\end{tabular}} \\ \hline
\multirow{3}{*}{YOLOv8n}          & pytorch         & 0.692                                                      & 0.39                                                          & 0.752              & 0.615           & \multirow{3}{*}{3006428}          & 48.2                                                               & 6894.8                                                            & 13.8                                                                & 6956.8                                                          \\
                                  & onnx            & 0.685                                                      & 0.385                                                         & 0.742              & 0.618           &                                   & 31.6                                                               & 3031.1                                                            & 21.1                                                                & 3083.8                                                          \\
                                  & ncnn            & 0.685                                                      & 0.385                                                         & 0.742              & 0.618           &                                   & 26.2                                                               & 2780                                                              & 13.7                                                                & 2819.9                                                          \\ \hline
\multirow{3}{*}{\textbf{YOLOv9t}} & pytorch         & 0.686                                                      & 0.385                                                         & 0.747              & 0.621           & \multirow{3}{*}{\textbf{1971564}} & 28.5                                                               & 5429.7                                                            & 11.4                                                                & 5469.6                                                          \\
                                  & onnx            & 0.687                                                      & 0.383                                                         & 0.746              & 0.621           &                                   & 24.7                                                               & 3154.3                                                            & 18.1                                                                & 3197.1                                                          \\
                                  & \textbf{ncnn}   & \textbf{0.687}                                             & \textbf{0.383}                                                & \textbf{0.746}     & \textbf{0.621}  &                                   & \textbf{16.9}                                                      & \textbf{1968.7}                                                   & \textbf{9.2}                                                        & \textbf{1994.8}                                                 \\ \hline
\multirow{2}{*}{YOLOv10n}         & pytorch         & 0.659                                                      & 0.367                                                         & 0.684              & 0.598           & \multirow{2}{*}{2695976}          & 37.3                                                               & 6004.8                                                            & 1.5                                                                 & 6043.6                                                          \\
                                  & onnx            & 0.657                                                      & 0.366                                                         & 0.695              & 0.597           &                                   & 20.3                                                               & 2307.1                                                            & 1.4                                                                 & 2328.8                                                          \\ \hline
\end{tabular}
}
\label{tab:model-perf-pi}
\end{table*}


\vspace{-0.2cm}
\subsection{Indoor Testing}
We test our developed system in a simulated environment inside a room to determine the functionality of our system. While the processing pipeline is running, we monitor the Raspberry Pi CPU and evaluate the feasibility of the pipeline in working conditions (low resource environment).

\subsubsection{Procedure of Experimentation}
\begin{figure}
\centering
  \begin{subfigure}[t]{.475\linewidth}
    \centering
    \includegraphics[width=\linewidth, height=0.55\textwidth]{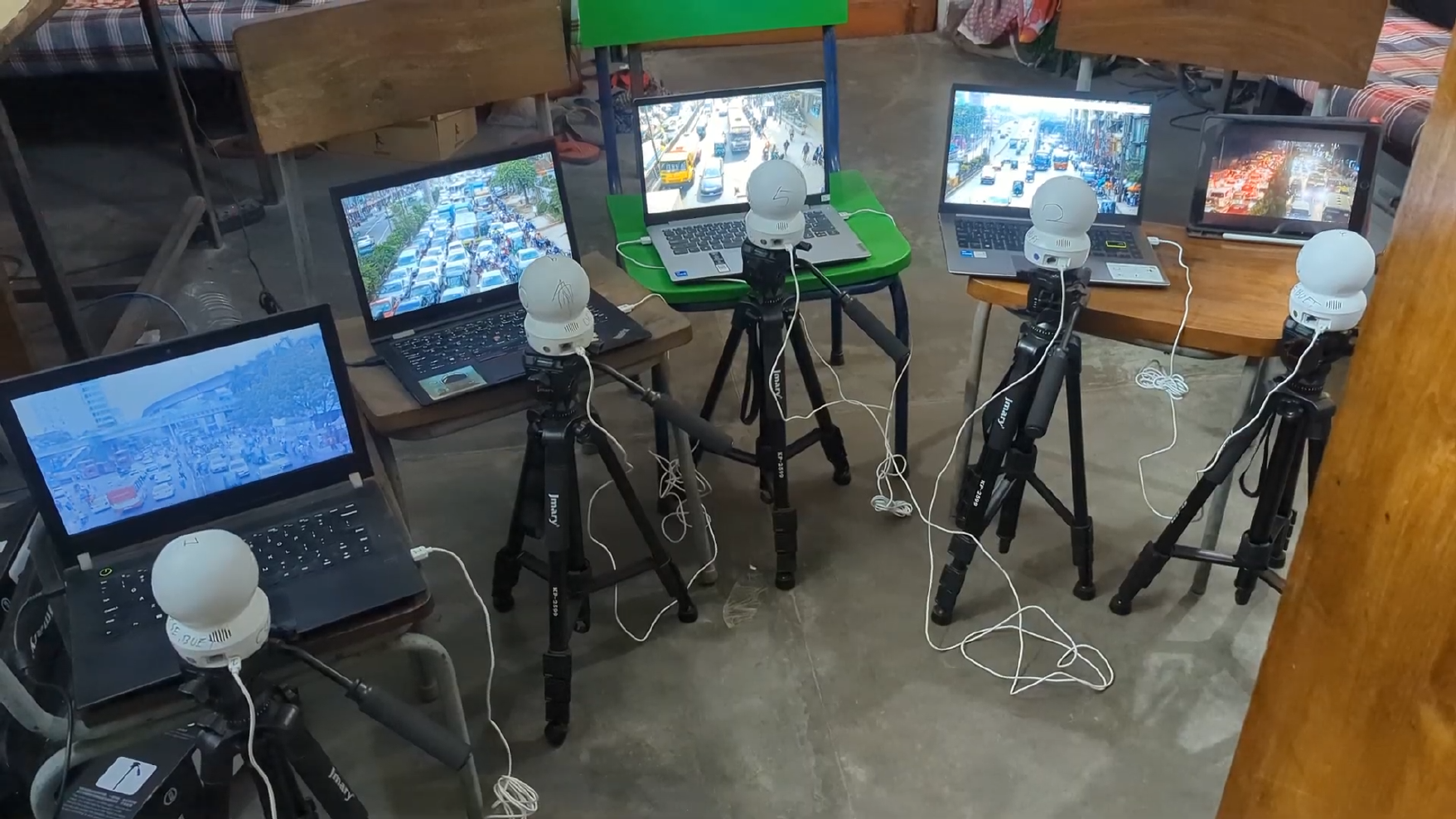}
    \label{fig:indoor-1}
  \end{subfigure}
  \begin{subfigure}[t]{.475\linewidth}
    \centering
    \includegraphics[width=\linewidth, height=0.55\textwidth]{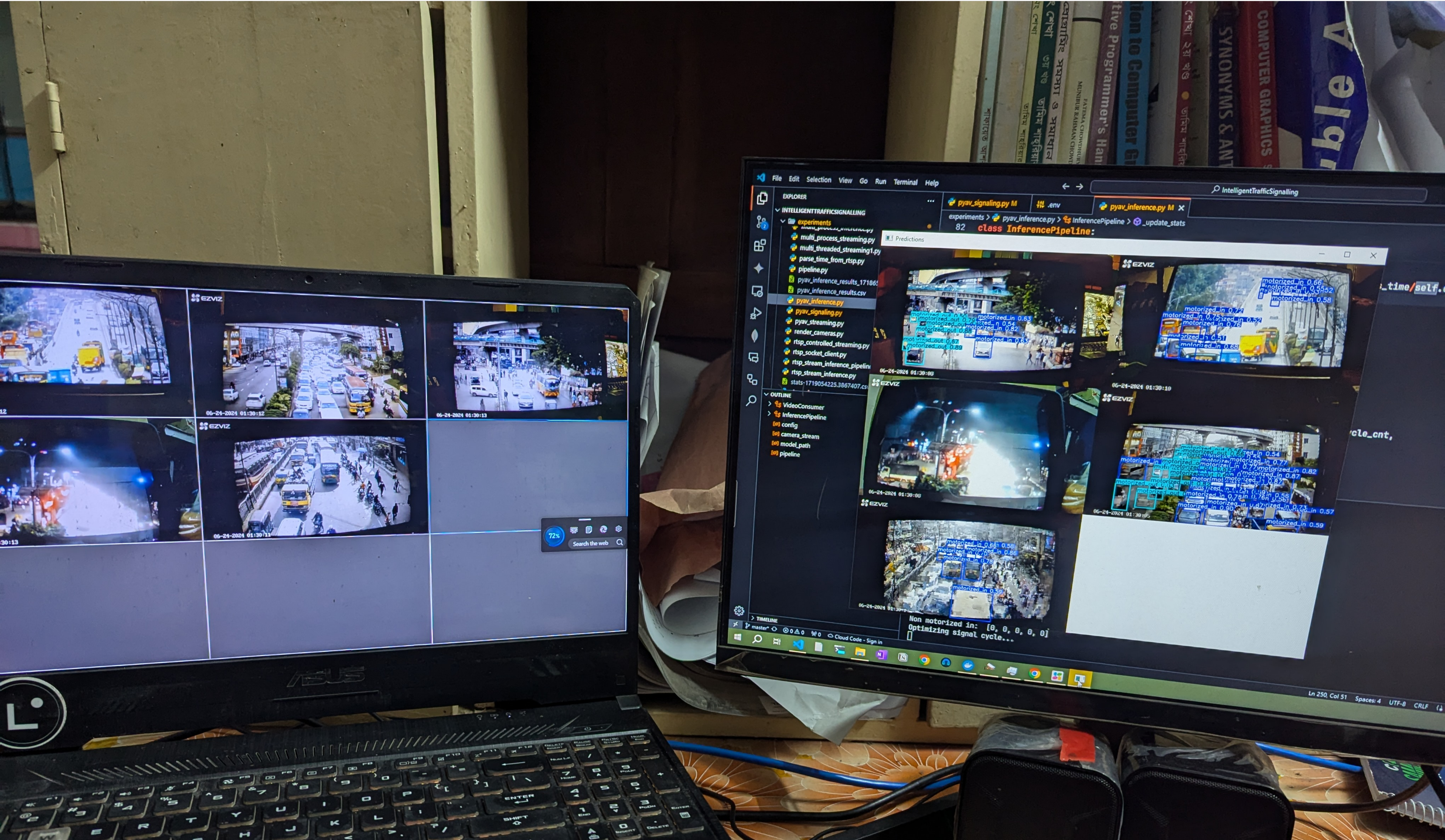}
    \label{fig:indoor-3}
  \end{subfigure}
  \vspace{-0.25cm}
\caption{Indoor testing of our system}
\label{fig:indoor-test}
\vspace{-0.55cm}
\end{figure}

The indoor testing of our system took place in a controlled simulated environment where IP cameras were positioned to capture five screens that displayed YouTube videos of Dhaka traffic. Our primary goal was to verify the functionality of the system and check the feasibility of the hardware implementation. Figure \ref{fig:indoor-test} shows snippets from the indoor test of our system.

\subsubsection{Evaluation Metrics}
We evaluate the feasibility of the implementation based on the following metrics:

\vspace{0.3cm}
\textit{CPU Utilization}: 
The percentage of the CPU's processing capacity used by the system while the processing pipeline is active. This metric is crucial for assessing the efficiency and workload management of the processing pipeline in a low resource environment. 

\vspace{0.3cm}
\textit{CPU Temperature}: 
The temperature of the CPU during the experiment is monitored to ensure that the Raspberry Pi remains within safe operational limits. Overheating could indicate inefficiencies or the need for better cooling mechanisms. 

\begin{figure*}[]
    \centering
    \includegraphics[width=0.7\linewidth]{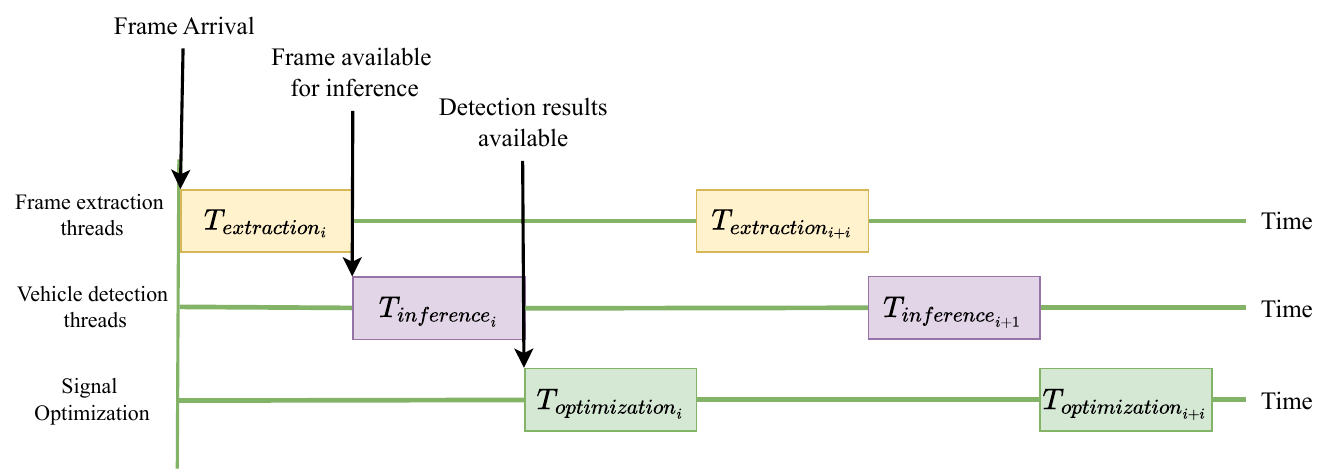}
    \vspace{-0.25cm}
    \caption{Breakdown of end-to-end latency in our developed system}
    \label{fig:exp:metric:timing}
    \vspace{-0.25cm}
\end{figure*}

\vspace{0.3cm}
\textit{End-to-end Latency}: 
Latency is vital for evaluating the real-time performance and responsiveness of the system. Figure \ref{fig:exp:metric:timing} depicts the breakdown of end-to-end latency in our processing pipeline, as described by equations  (\ref{eq:extraction_time}), (\ref{eq:inference_time}), and (\ref{eq:latency_individual}). It starts with the moment a video frame enters the system ("Frame Arrival"). There's a delay before the frame is prepared for processing by the object detection model. This delay, captured by equation  (\ref{eq:extraction_time}), includes the time taken to extract the frame from the video stream using the pyav library with multithreading. Once prepared, the object detection model analyzes the frame, and the time taken for this analysis is reflected in the latency between "Frame available for inference" and "Detection results available" (refer to equation (\ref{eq:inference_time})). Finally, the NSGA-II algorithm takes some time to optimize the signal timing based on the detected traffic.

The sum of these individual delays for each cycle (denoted by $T_{latency_i}$ in equation (\ref{eq:latency_individual})) represents the total latency experienced in a single cycle. Equation (\ref{eq:latency_total}) depicts the calculation of the overall end-to-end latency ($T_{latency}$) by averaging the individual latency across multiple cycles (represented by N). Minimizing this overall latency is critical because it allows the system to react more quickly to changes in traffic flow, potentially leading to reduced congestion.
\vspace{-0.2cm}
\begin{equation}
    \label{eq:extraction_time}
    T_{extraction_i} = \frac{\displaystyle\sum_{j=0}^{n_f} D_{extraction_j}}{n_f}
\end{equation}
\vspace{-0.2cm}
\begin{equation}
    \label{eq:inference_time}
    T_{inference_i} = \frac{\displaystyle\sum_{j=0}^{n_f} D_{inference_j}}{n_f}
\end{equation}
\vspace{-0.2cm}
\begin{equation}
    \label{eq:latency_individual}
    T_{latency_i} = T_{extraction_i} + T_{inference_i} + T_{optimization_i}
\end{equation}
\vspace{-0.2cm}
\begin{equation}
    \label{eq:latency_total}
    T_{latency} = \frac{\displaystyle\sum_{i=0}^{N} T_{latency_i}}{N}
\end{equation}
\vspace{-0.2cm}
\subsubsection{Evaluation based on Indoor Testing}

\begin{figure}[!t]
    \centering
    
    \begin{subfigure}[t]{0.9\linewidth}
        \centering
        \includegraphics[width=\linewidth]{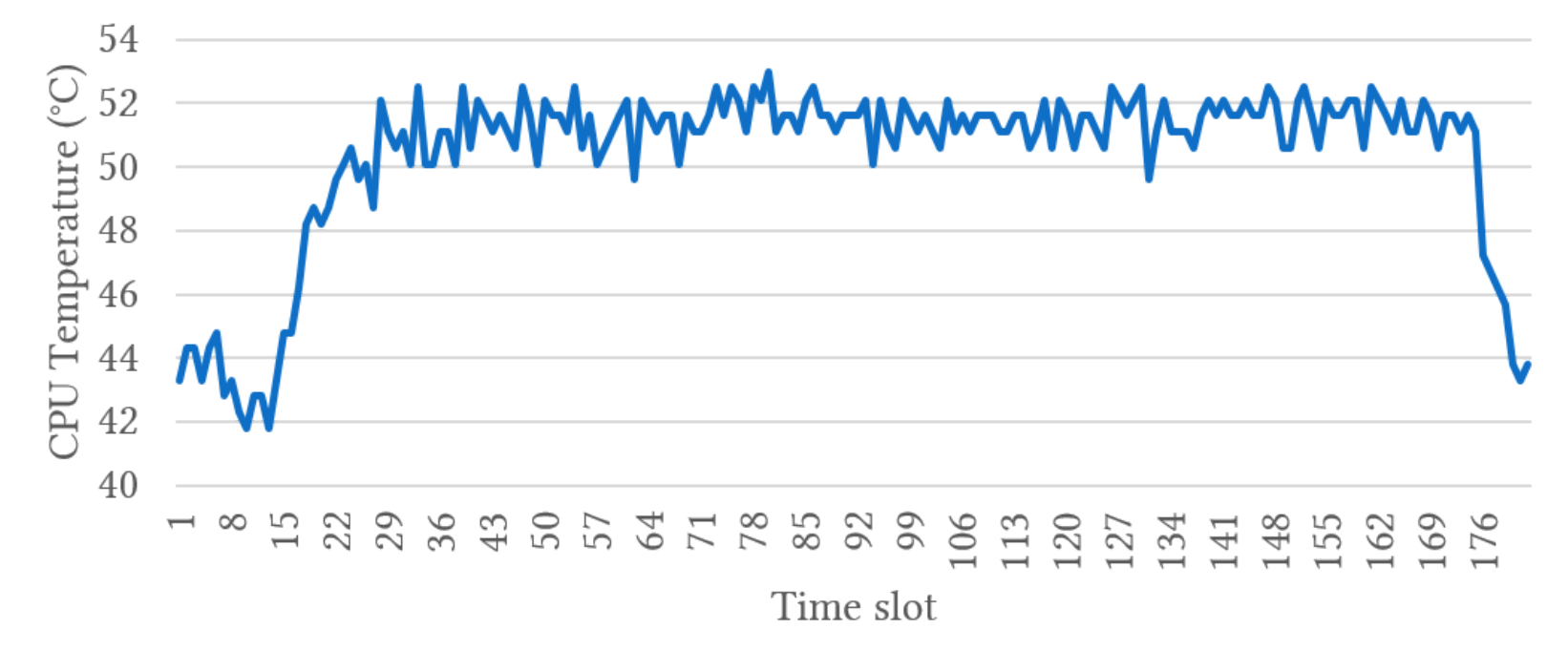}
        \caption{CPU temperature}
        \label{fig:stats:indoor-temp}
    \end{subfigure}
    \begin{subfigure}[t]{0.9\linewidth}
        \centering
        \includegraphics[width=\linewidth]{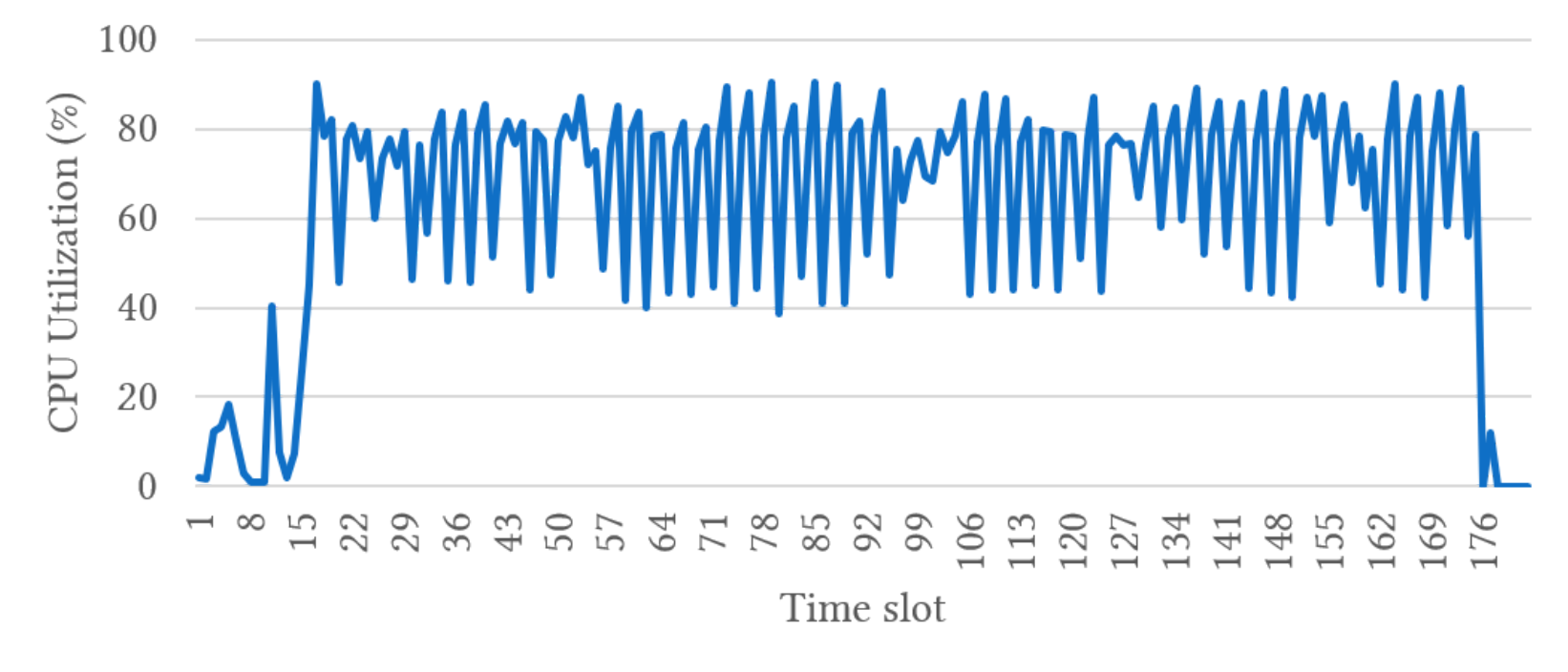}
        \caption{CPU Utilization}
        \label{fig:stats:indoor-cpu-usage}
    \end{subfigure}
    \vspace{-0.25cm}
    \caption{Raspberry Pi CPU Statistics (Indoor Testing)}
    \label{fig:met:pi-stats-indoor}
    \vspace{-0.55cm}
\end{figure}

Figure \ref{fig:met:pi-stats-indoor} shows the change of the aforementioned metrics while the system is active during indoor testing. The Raspberry Pi does not heat up a lot during indoor testing. The operating temperature for a Raspberry Pi is between 0°C and 85°C, and in Figure \ref{fig:stats:indoor-temp} we see the highest temperature recorded during the testing is 53\textdegree C.

In Figure \ref{fig:stats:indoor-cpu-usage}, we notice the CPU utilization is almost constant while the pipeline is running. Since it does not reach a critical usage, the system is viable in terms of CPU utilization.

\begin{figure}[!t]
    \centering
    \includegraphics[width=0.8\linewidth]{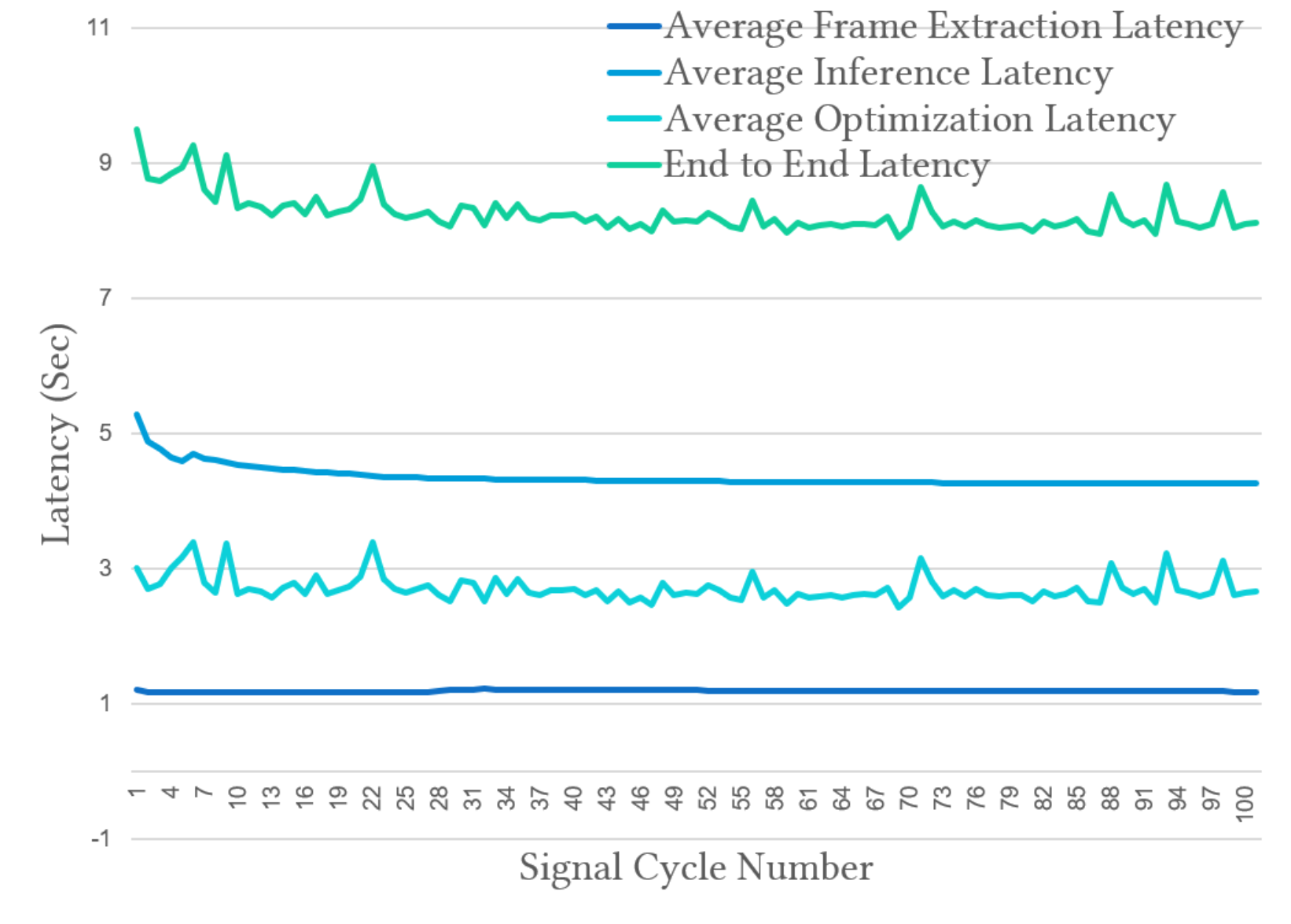}
    \vspace{-0.25cm}
    \caption{Latency during indoor testing.}
    \label{fig:met:latency-indoor}
    \vspace{-0.55cm}
\end{figure}

The end-to-end latency of the system during the test remains constant as well, as seen in Figure \ref{fig:met:latency-indoor}. It is important because if the graph showed that the latency increases with running time, the system would be accumulating lag, and thus become impracticable to use. Too much latency would cause the system to be impracticable as well, since the scenario of traffic may change in the time the signal cycle is processed. The latency reaches around 8 seconds, which falls under a reasonable threshold in the context of traffic signaling.

\vspace{-0.2cm}
\subsection{On-road Testing}
On-road testing is the most significant and interesting part of the study. We deploy the system in an intersection to understand if the signal cycles scheduled by the system are efficient in minimizing congestion. Since the output of our system does not involve traffic lights in the testbed, we take help from the stationed traffic police at the intersection. During the time the system is operating and processing pipeline is running, they guide the traffic according to the output signal cycles of our system. We log necessary information during the signaling for further analysis. It is to be noted that, we have taken approval from Ethics Committee (affiliated with the university of the corresponding author) prior to conducting the on-road data collection.

The location of our on-road testing is the Palashi intersection in Dhaka, Bangladesh. It is a five-way intersection with heterogeneous traffic throughout the day. The links connected to the intersection are: 
(a) Link 1: Towards Fuller Road, 
(b) Link 2: Towards Bakshibazar, 
(c) Link 3: Towards Azimpur, 
(d) Link 4: Towards Nilkhet, 
(e) Link 5: Towards Dhakeshwari.

\begin{figure}
    \centering
    \includegraphics[width=0.5\linewidth]{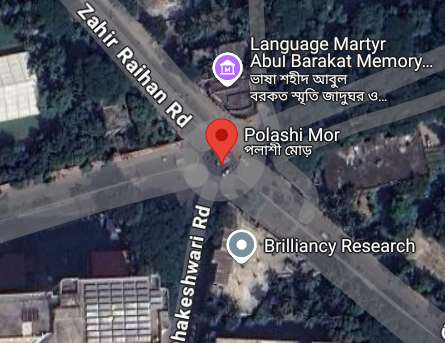}
    \vspace{-0.25cm}
    \caption{Google Maps snippet of the on-road testing location (Palashi intersection, Dhaka)}
    \vspace{-0.25cm}
    \label{fig:exp:on-road:location}
\end{figure}


\begin{figure}
\centering
  \begin{subfigure}[t]{.3\linewidth}
    \centering
    \includegraphics[width=\linewidth, height=0.55\textwidth]{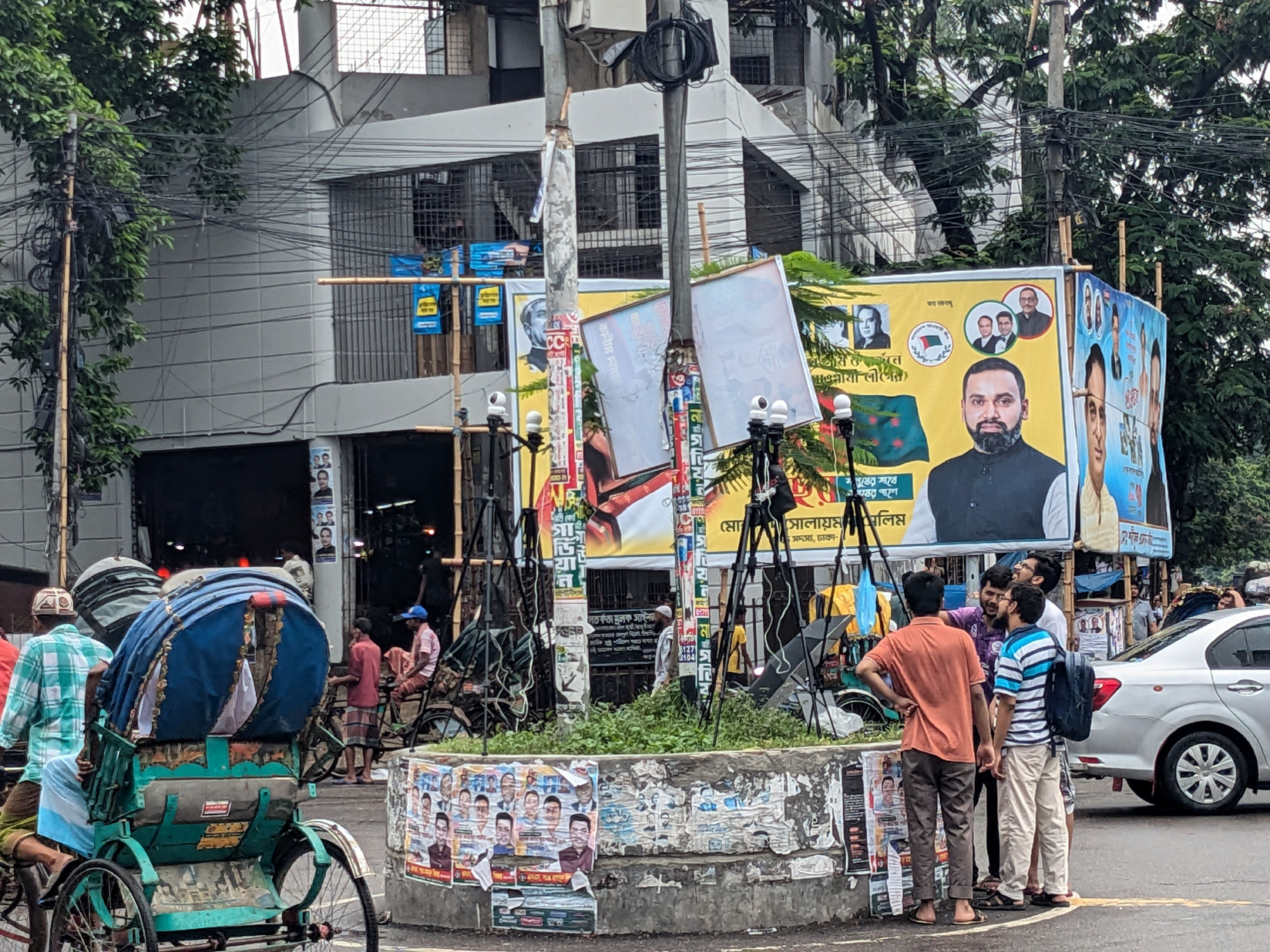}
    \label{fig:nht1}
  \end{subfigure}
  \begin{subfigure}[t]{.3\linewidth}
    \centering
    \includegraphics[width=\linewidth, height=0.55\textwidth]{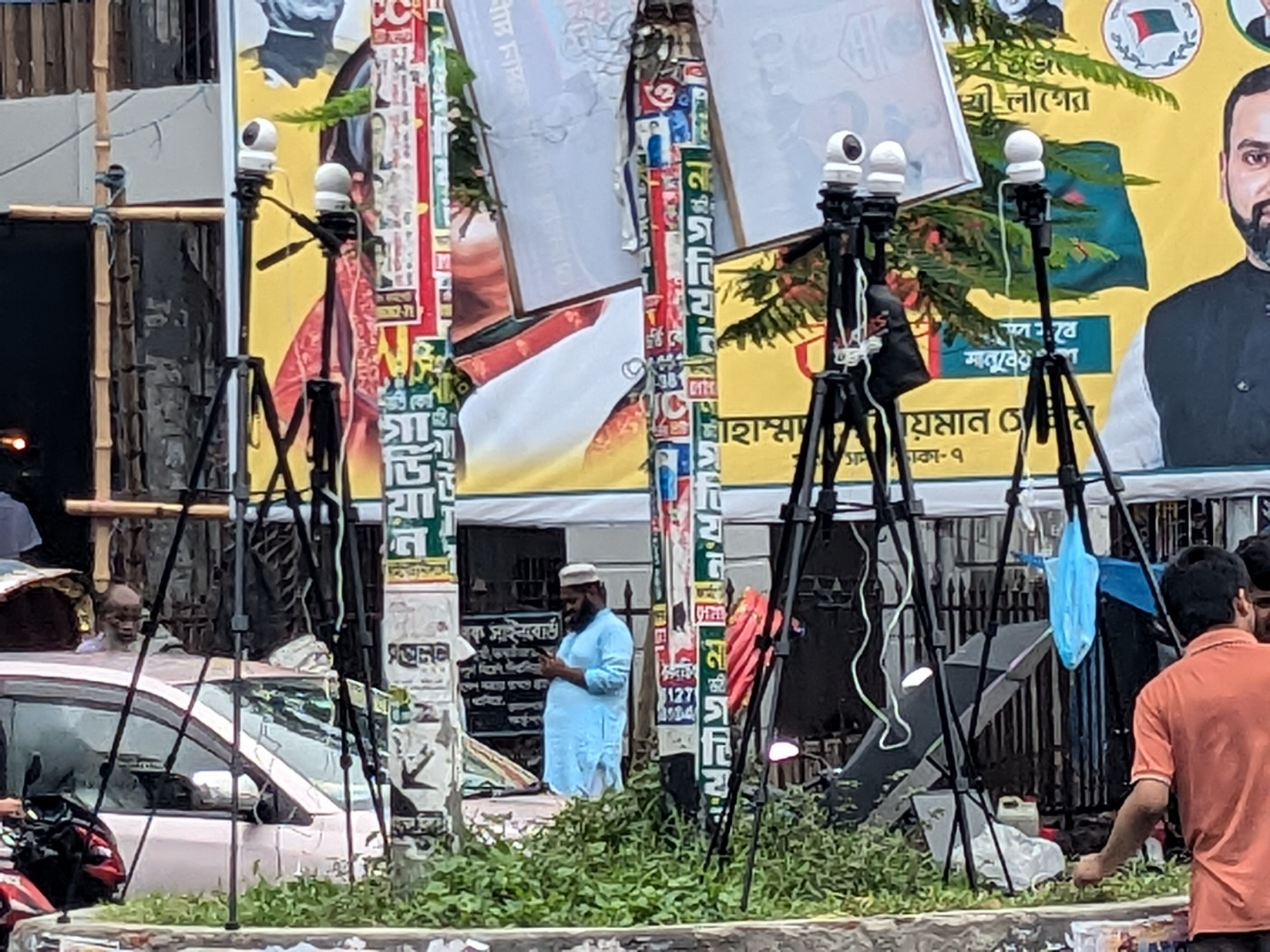}
    \label{fig:nht2}
  \end{subfigure}
  \begin{subfigure}[t]{.3\linewidth}
    \centering
    \includegraphics[width=\linewidth, height=0.55\textwidth]{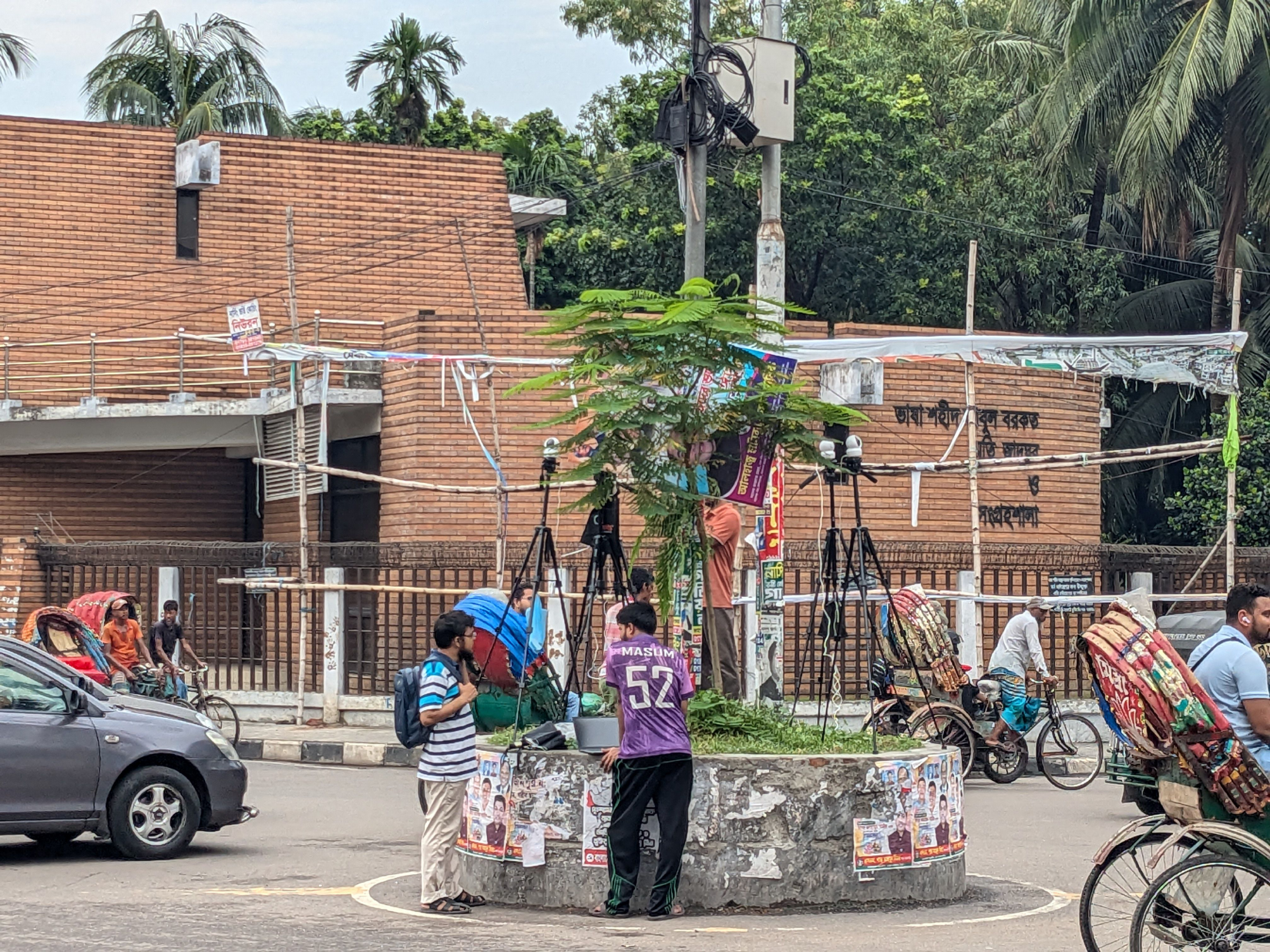}
    \label{fig:nht2}
  \end{subfigure}
  \vspace{-0.4cm}
\caption{Conducting experiments on-road at Palashi intersection, Dhaka}
\label{fig:on-road-on}
\vspace{-0.5cm}
\end{figure}

Based on our gathered experiences and findings, we conduct our experiments in five different phases. We discuss the setups and observations in this section, while the relevant findings are discussed in the following chapter.

\subsubsection{Evaluation Metrics}
We evaluate the system based on the metrics mentioned in indoor testing (CPU utilization, CPU temperature and end-to-end latency) as well as new metrics to compare the signal cycle scheduling to the manual signaling by the traffic police. 
The new metrics for on-road evaluation are maximum waiting vehicles in a link, and average waiting vehicles in per link. 
These metrics denote the degree of congestion control by each signaling technique.

\textit{Maximum Waiting Vehicles in a link}:
This metric simply denotes the highest number of vehicles waiting in a link at a certain time slot. It denotes the maximum extent of traffic congestion caused by the signaling.

\textit{Average Waiting Vehicles per link}:
The mean of the number of vehicles at each link is used to denote the overall effect of the signaling on the traffic condition.

We ask the stationed police to control the traffic according to the system proposed signal cycle, and afterwards, we recorded data relevant to the comparison (vehicles waiting in the five links) while they controlled the traffic.

\subsubsection{Phases of Experimentation}

We can divide our on-road experiments into five different phases chronologically. We took some observations from the previous phase into consideration for the next phases. 

\textbf{Phase 0} - \textit{Day of the week: Friday, Time: 6:00 PM (Evening)}

\begin{figure}[!t]
    \centering
    
    \begin{subfigure}[t]{0.9\linewidth}
        \centering
        \includegraphics[width=\linewidth]{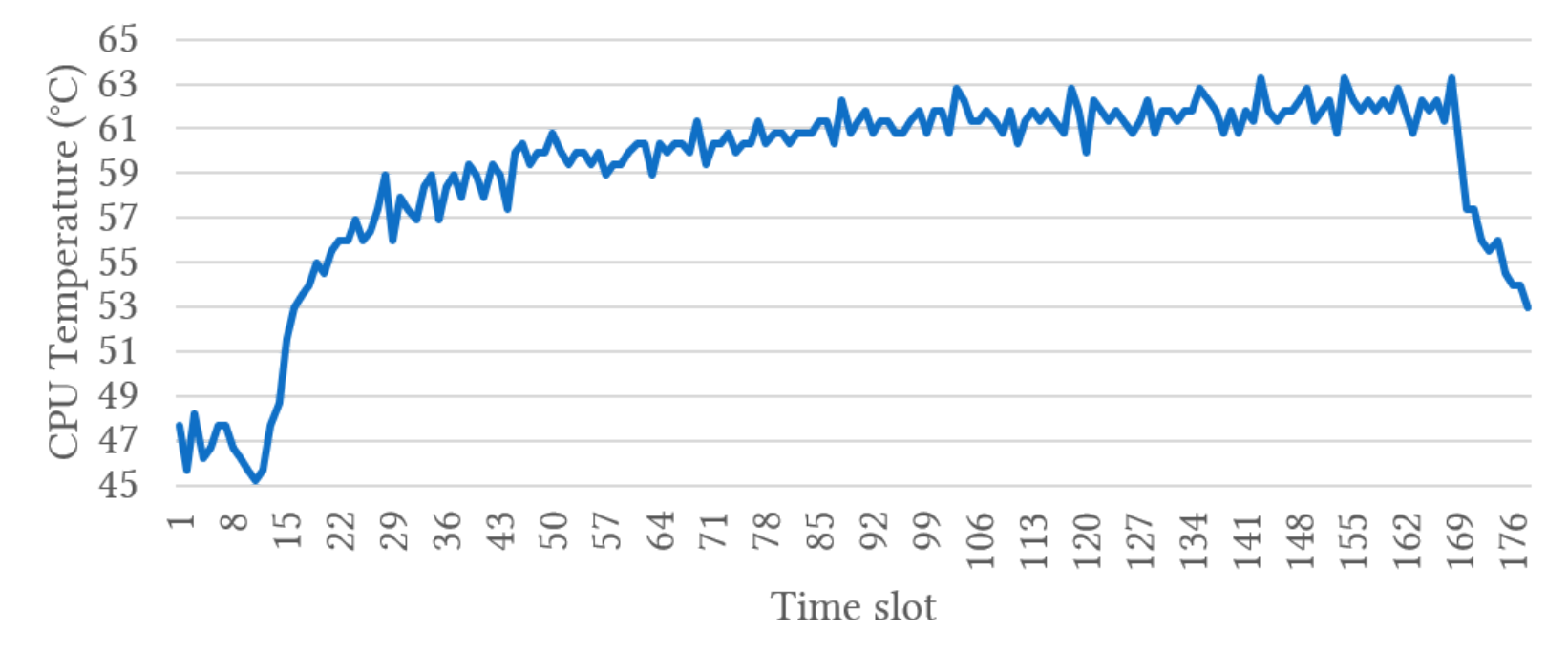}
       \caption{CPU temperature}
        \label{fig:stats:outdoor-temp}
    \end{subfigure}
    \begin{subfigure}[t]{0.9\linewidth}
        \centering
        \includegraphics[width=\linewidth]{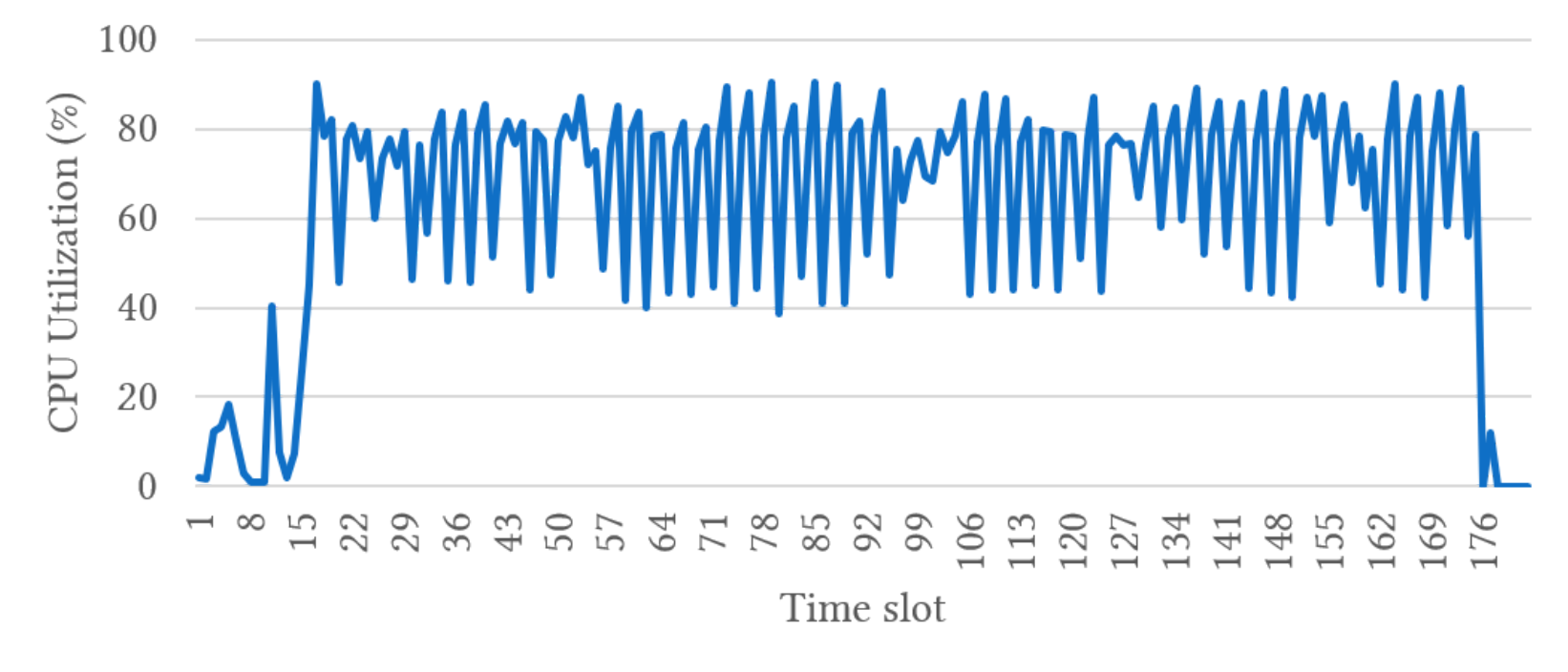}
        \caption{CPU Utilization}
        \label{fig:stats:outdoor-cpu-usage}
    \end{subfigure}
    \vspace{-0.25cm}
    \caption{Raspberry Pi CPU Statistics (On-road)}
    \label{fig:met:pi-stats-on-road}
    \vspace{-0.55cm}
\end{figure}

The first phase of the experiment was to understand the feasibility of the location and setting up. We set up the cameras on tripods, positioning them to face each link atop the roundabout. Following this, we checked the connectivity of the cameras to the local network via a mobile hotspot, ensuring that they were properly linked with the Raspberry Pi. Unfortunately, the pipeline could not operate during the experiment.

The experiment faced challenges that hindered the operation of the system. Headlights flashing directly at the cameras resulted in minimal vehicle detection. Additionally, the cameras were positioned low, which not only impeded unobstructed capturing of the respective links but also increased susceptibility to glare from headlights. The amount of traffic observed during the test was insufficient for proper experimentation, indicating that future tests need to be conducted during busier hours to yield better results. 

\textbf{Phase 1} - \textit{Day of week: Wednesday, Time: 1:00 PM (Afternoon)}

The setup for this phase was the same as the previous phase, Phase 0. The system was operative in this phase, and we collected CPU data which is shown in Figure \ref{fig:met:pi-stats-on-road}. We can see that the Raspberry Pi CPU gets to a higher temperature during the on-road testing than during indoor testing, but not crossing 85\textdegree Celsius, which is the limit for a Pi CPU.

Although the processing pipeline was functional, signaling could not be done during Phase 1. Additionally, the distance from Link 5 to the respective camera was large, affecting the inference results.

\textbf{Phase 2} - \textit{Day of week: Thursday, Time: 10:30 AM}

During a busier traffic period with a larger volume of vehicles than in previous experiments, we initially attempted a different setup for Link 5 by positioning the camera at a separate location. However, this setup failed due to a poor connection to the local network, leading us to revert to the original configuration. During this phase, signal scheduling output by the system was compared to that conducted by the traffic police.

The signal scheduling was interrupted by an ambulance movement, prompting the green time to be extended for the link with the ambulance beyond the scheduled duration. System loss was observed due to traffic guidance during signaling.

\textbf{Phase 3} - \textit{Day of week: Thursday, Time: 4:00 PM (Afternoon)}

During a period of busy traffic with a large volume of vehicles, the signal scheduling was compared to the signal scheduling conducted by traffic police. The setup remained the same as Phase 0.

In this phase, the signal scheduling was briefly interrupted due to VIP movement but resumed shortly thereafter. System loss was observed due to traffic guidance during signaling. Light glare affected the cameras facing Link 3 and 5. Additionally, crossing pedestrians caused intermittent interruptions in the traffic flow during the experiment.

\textbf{Phase 4} - \textit{Day of week: Friday, Time: 4:30 PM (Afternoon)}

During a period of moderate traffic volume, signal scheduling was performed with specific considerations. To account for system loss due to traffic police guidance, eight seconds were added to each green time, four seconds before and after the green time provided by the signal cycle. Additionally, to accommodate ambulance or any emergency vehicle during the time of testing, the cycle was not kept in order. For example, if an ambulance was approaching the intersection through Link 2, the traffic police was instructed to allow Link 2 to open instead of the link which were in order.

In this phase, the signaling system demonstrated improved performance and sustained operation for a longer duration compared to previous trials. Additionally, sunlight glare affecting the camera facing Link 3 was observed during the experiment.





\vspace{-0.4cm}
\section{Discussion} \label{sec:discussion}
After testing and evaluating our proposed system, we discuss the factors of consideration in deployment and the limitations we faced in our study.
\vspace{-0.3cm}
\subsection{Deployment Considerations}

\subsubsection*{System Performance}
Our experiments indicate that key performance metrics such as end-to-end latency, CPU utilization, and CPU temperature remain nearly unchanged throughout the running time of the signal scheduling pipeline. This consistency suggests that our system is both efficient and robust, capable of maintaining optimal performance without imposing significant computational overhead. The consistency highlights potential for scalability and ability to handle real-time traffic signal adjustments without degradation in processing speed or system stability in low resource environments. More importantly than the other factors, the end-to-end latency reported by the system is low enough to generate a near-instant traffic signaling response. It shows that the system is viable while concurrently processing video streams of up to five roads. The evaluation of this metric vouches for the scalability of the system.

\subsubsection*{Camera Position and Shielding}
One of the key factors in real-time traffic signaling performance is the accuracy of the vehicle detection model, which in turn depends on the extracted frames from the camera streams. While our system demonstrates effectiveness in traffic signaling real-time, the point of view of the cameras require further modifications for better results. Interference due to crossing vehicles and pedestrians affect the count of the incoming vehicles from a link. Glares from sunlight or headlights severely impact the performance of the model as well. It is required to improve the placement of the camera along with shielding techniques for sunlight and averse weather conditions as well. A higher point of installment should improve the results of the model and subsequently the system in total.


\subsubsection*{Manual Traffic Guidance and its Variability}
After a signal is scheduled, the time taken for the traffic police to guide the traffic varies significantly. Unexpected events, traffic density and other factors influence the time taken by the traffic personnel. This delay would not persist after the complete infrastructural deployment of our system, but for testing, an integrated approach is required. Other tools to assist the traffic police may also diminish the delay due to traffic guidance.


\subsubsection*{Emergency Vehicle Movement}
Due to the on-road testing location being very close to some hospitals in Dhaka city, we faced the problem of accommodating ambulances in our scheduled signal cycles. In Phase 4 of our on-road testing, we adapt to the situation by reordering the cycle once an ambulance is spotted waiting at a link to ensure smooth movement of such transport.

\vspace{-0.2cm}
\subsection{Limitations}
\subsubsection*{Consideration for Emergency Vehicles}
Emergency vehicles and similar protocols require faster movement on road than regular traffic, but our system does not accommodate for their movement. Advancements are required in this aspect for the system to be operable in the intersections of Dhaka city.

\subsubsection*{Infrastructural Support}
The IP cameras on tripods provide a viewpoint from a lower height than optimal, causing visual interference in the streams due to pedestrians, glares from sunlight and other factors mentioned before. There are cameras installed in the intersections of Dhaka city, which are inoperative or only used for monitoring the traffic. Such cameras are also present in our testing location, where the cameras in an optimal height for vehicle detection in each lane. The infrastructural support of accessing cameras installed at an optimal height for our context or simply access to install cameras at that height would improve the effects of the working system.

\subsubsection*{Full Implementation including Signal Lights}
Traffic signaling with manual guidance adds some overhead in our pipeline, causing delays to reflect the scheduled signal cycle on road. The integration of the proposed system with traffic lights would unlock the potential of our system in eradicating congestion and relevant traffic issues in Dhaka city.

\vspace{-0.25cm}
\section{Conclusion} \label{sec:conclusion}
Decentralized traffic-responsive signal systems present a promising solution to the traffic congestion in the cities of developing countries, such as Dhaka, Bangladesh. Our proposed architecture serves as a foundational framework for such systems, specifically designed to address the unique challenges posed by non-lane-based heterogeneous traffic prevalent in these regions. By leveraging advanced algorithms and real-time data, our system dynamically adjusts signal timings to optimize traffic flow, reduce delays, and enhance overall road efficiency.

Throughout this study, we have demonstrated the feasibility and effectiveness of our architecture through comprehensive simulations and real-world implementations in Dhaka city. The results indicate that our system meets the performance of prevalent traffic signal scheduling methods. By offering comparable or improved outcomes in terms of reduced congestion and enhanced vehicle throughput, our approach significantly contributes to a more efficient and reliable urban transportation network.

Moreover, the decentralized nature of our system ensures scalability and adaptability, making it suitable for deployment in various urban or semi-urban settings with diverse traffic patterns. To the best of our knowledge, our proposed system architecture is the very first implementation of IoT-based adaptive signaling system in Dhaka city.

\ifCLASSOPTIONcaptionsoff
  \newpage
\fi



\bibliographystyle{IEEEtran}
\bibliography{sample-base}

\begin{thebibliography}{10}
\providecommand{\url}[1]{#1}
\csname url@samestyle\endcsname
\providecommand{\newblock}{\relax}
\providecommand{\bibinfo}[2]{#2}
\providecommand{\BIBentrySTDinterwordspacing}{\spaceskip=0pt\relax}
\providecommand{\BIBentryALTinterwordstretchfactor}{4}
\providecommand{\BIBentryALTinterwordspacing}{\spaceskip=\fontdimen2\font plus
\BIBentryALTinterwordstretchfactor\fontdimen3\font minus \fontdimen4\font\relax}
\providecommand{\BIBforeignlanguage}[2]{{%
\expandafter\ifx\csname l@#1\endcsname\relax
\typeout{** WARNING: IEEEtran.bst: No hyphenation pattern has been}%
\typeout{** loaded for the language `#1'. Using the pattern for}%
\typeout{** the default language instead.}%
\else
\language=\csname l@#1\endcsname
\fi
#2}}
\providecommand{\BIBdecl}{\relax}
\BIBdecl

\bibitem{chen2020future}
B.~Chen, D.~Sun, J.~Zhou, W.~Wong, and Z.~Ding, ``A future intelligent traffic system with mixed autonomous vehicles and human-driven vehicles,'' \emph{Information Sciences}, vol. 529, pp. 59--72, 2020.

\bibitem{yogheshwaran2020iot}
M.~Yogheshwaran, D.~Praveenkumar, S.~Pravin, P.~Manikandan, and D.~S. Saravanan, ``Iot based intelligent traffic control system,'' \emph{International Journal of Engineering Technology Research \& Management}, vol.~4, no.~4, pp. 59--63, 2020.

\bibitem{liu2020thresholds}
Y.~Liu, C.~Yang, and Q.~Sun, ``Thresholds based image extraction schemes in big data environment in intelligent traffic management,'' \emph{IEEE transactions on intelligent transportation systems}, 2020.

\bibitem{beg2021uav}
A.~Beg, A.~R. Qureshi, T.~Sheltami, and A.~Yasar, ``Uav-enabled intelligent traffic policing and emergency response handling system for the smart city,'' \emph{Personal and Ubiquitous Computing}, vol.~25, no.~1, pp. 33--50, 2021.

\bibitem{zhou2021construction}
X.~Zhou, ``Construction and application of urban intelligent traffic control system based on cloud computing,'' in \emph{Journal of Physics: Conference Series}, vol. 1952, no.~4.\hskip 1em plus 0.5em minus 0.4em\relax IOP Publishing, 2021, p. 042006.

\bibitem{mrahman23}
M.~Rahaman, A.~Mahir Sultan~Rumi, M.~Shihabul~Islam, T.~Reza~Toha, M.~Masum~Mushfiq, M.~Sohel~Rahman, M.~Ali~Nayeem, N.~Abdulrahman Al-Nabhan, and A.~B.~M. Alim Al~Islam, ``Toward devising a multi-objective traffic signal scheduling approach for non-lane-based heterogeneous traffic,'' \emph{IEEE Access}, vol.~13, pp. 172\,598--172\,616, 2025.

\bibitem{nsgaii}
K.~Deb, A.~Pratap, S.~Agarwal, and T.~Meyarivan, ``A fast and elitist multiobjective genetic algorithm: Nsga-ii,'' \emph{IEEE Transactions on Evolutionary Computation}, vol.~6, no.~2, pp. 182--197, 2002.

\bibitem{saha2020automated}
S.~Saha, ``Automated traffic law enforcement system: A feasibility study for the congested cities of developing countries: Automated traffic law enforcement system: A feasibility study for the congested cities of developing countries,'' \emph{International Journal of Innovative Technology and Interdisciplinary Sciences}, vol.~3, no.~1, pp. 346--363, 2020.

\bibitem{tajalli2020network}
M.~Tajalli, M.~Mehrabipour, and A.~Hajbabaie, ``Network-level coordinated speed optimization and traffic light control for connected and automated vehicles,'' \emph{IEEE Transactions on Intelligent Transportation Systems}, 2020.

\bibitem{al2019smart}
M.~Al-qutwani and X.~Wang, ``Smart traffic lights over vehicular named data networking,'' \emph{Information}, vol.~10, no.~3, p.~83, 2019.

\bibitem{waddell2020characterizing}
J.~M. Waddell, S.~M. Remias, and J.~N. Kirsch, ``Characterizing traffic-signal performance and corridor reliability using crowd-sourced probe vehicle trajectories,'' \emph{Journal of Transportation Engineering, Part A: Systems}, vol. 146, no.~7, p. 04020053, 2020.

\bibitem{yao2020reducing}
Z.~Yao, B.~Zhao, T.~Yuan, H.~Jiang, and Y.~Jiang, ``Reducing gasoline consumption in mixed connected automated vehicles environment: A joint optimization framework for traffic signals and vehicle trajectory,'' \emph{Journal of Cleaner Production}, vol. 265, p. 121836, 2020.

\bibitem{chen2021rhythmic}
X.~Chen, M.~Li, X.~Lin, Y.~Yin, and F.~He, ``Rhythmic control of automated traffic—part i: Concept and properties at isolated intersections,'' \emph{Transportation Science}, vol.~55, no.~5, pp. 969--987, 2021.

\bibitem{almawgani2018design}
A.~Almawgani, ``Design of real time smart traffic light control system,'' in \emph{Proceedings of ISER 108th International Conference, Mecca, Saudi Arabia}, 2018, pp. 51--55.

\bibitem{zhang2020}
R.~Zhang, A.~Ishikawa, W.~Wang, B.~Striner, and O.~Tonguz, ``Using reinforcement learning with partial vehicle detection for intelligent traffic signal control,'' \emph{IEEE Transactions on Intelligent Transportation Systems}, vol.~PP, pp. 1--12, 03 2020.

\bibitem{mandal2020}
V.~Mandal, A.~R. Mussah, P.~Jin, and Y.~Adu-Gyamfi, ``Artificial intelligence-enabled traffic monitoring system,'' \emph{Sustainability}, vol.~12, no.~21, 2020.

\bibitem{joo2022}
\BIBentryALTinterwordspacing
H.~Joo and Y.~Lim, ``Intelligent traffic signal phase distribution system using deep q-network,'' \emph{Applied Sciences}, vol.~12, no.~1, 2022. [Online]. Available: \url{https://www.mdpi.com/2076-3417/12/1/425}
\BIBentrySTDinterwordspacing

\bibitem{leal2023}
S.~Leal and P.~Almeida, ``Traffic light optimization using non-dominated sorting genetic algorithm (nsga2),'' \emph{Scientific Reports}, vol.~13, 09 2023.

\bibitem{rai2023}
\BIBentryALTinterwordspacing
S.~C. Rai, S.~P. Nayak, B.~Acharya, V.~C. Gerogiannis, A.~Kanavos, and T.~Panagiotakopoulos, ``Itss: An intelligent traffic signaling system based on an iot infrastructure,'' \emph{Electronics}, vol.~12, no.~5, 2023. [Online]. Available: \url{https://www.mdpi.com/2079-9292/12/5/1177}
\BIBentrySTDinterwordspacing

\bibitem{jahan2020analyzing}
M.~I. Jahan, A.~A.~B. Mazumdar, M.~Hadiuzzaman, S.~M. Mashrur, and M.~N. Murshed, ``Analyzing service quality of pedestrian sidewalks under mixed traffic condition considering latent variables,'' \emph{Journal of Urban Planning and Development}, vol. 146, no.~2, p. 04020011, 2020.

\bibitem{maungmai2019vehicle}
W.~Maungmai and C.~Nuthong, ``Vehicle classification with deep learning,'' in \emph{2019 IEEE 4th international conference on computer and communication systems (ICCCS)}.\hskip 1em plus 0.5em minus 0.4em\relax IEEE, 2019, pp. 294--298.

\bibitem{liu2018urban}
Z.~Liu, Z.~Li, K.~Wu, and M.~Li, ``Urban traffic prediction from mobility data using deep learning,'' \emph{Ieee network}, vol.~32, no.~4, pp. 40--46, 2018.

\bibitem{tan2017optimization}
M.~K. Tan, H.~S.~E. Chuo, R.~K.~Y. Chin, K.~B. Yeo, and K.~T.~K. Teo, ``Optimization of traffic network signal timing using decentralized genetic algorithm,'' in \emph{2017 IEEE 2nd International Conference on Automatic Control and Intelligent Systems (I2CACIS)}.\hskip 1em plus 0.5em minus 0.4em\relax IEEE, 2017, pp. 62--67.

\bibitem{nguyen2016discovering}
H.~Nguyen, W.~Liu, and F.~Chen, ``Discovering congestion propagation patterns in spatio-temporal traffic data,'' \emph{IEEE Transactions on Big Data}, vol.~3, no.~2, pp. 169--180, 2016.

\bibitem{dhakanet}
T.~R. Toha, M.~Rahaman, S.~I. Salim, M.~Hossain, A.~M. Sadri, and A.~B. M.~A. Al~Islam, ``Dhakanet: Unstructured vehicle detection using limited computational resources,'' in \emph{2021 IEEE International Conference on Data Mining (ICDM)}, 2021, pp. 1367--1372.

\bibitem{yolov5}
\BIBentryALTinterwordspacing
G.~Jocher, ``Ultralytics yolov5,'' 2020. [Online]. Available: \url{https://github.com/ultralytics/yolov5}
\BIBentrySTDinterwordspacing

\bibitem{yolov8_ultralytics}
\BIBentryALTinterwordspacing
G.~Jocher, A.~Chaurasia, and J.~Qiu, ``{Ultralytics YOLO},'' Jan. 2023. [Online]. Available: \url{https://github.com/ultralytics/ultralytics}
\BIBentrySTDinterwordspacing

\bibitem{wang2024yolov9}
C.-Y. Wang and H.-Y.~M. Liao, ``{YOLOv9}: Learning what you want to learn using programmable gradient information,'' 2024.

\bibitem{THU-MIGyolov10}
L.~L. Ao~Wang, Hui~Chen, ``Yolov10: Real-time end-to-end object detection,'' \emph{arXiv preprint arXiv:2405.14458}, 2024.

\bibitem{lv2023detrs}
W.~Lv, S.~Xu, Y.~Zhao, G.~Wang, J.~Wei, C.~Cui, Y.~Du, Q.~Dang, and Y.~Liu, ``Detrs beat yolos on real-time object detection,'' 2023.

\bibitem{ncnn}
\BIBentryALTinterwordspacing
H.~Ni and T.~ncnn contributors, ``ncnn,'' if you use this software, please cite it using the metadata from this file. [Online]. Available: \url{https://github.com/Tencent/ncnn}
\BIBentrySTDinterwordspacing

\end{thebibliography}
%


%








\end{document}